\documentclass[twocolumn,preprint]{elsarticle}
\usepackage{mypackages}
\journal{Journal of Computational Physics}

\begin{document}

\twocolumn[{
\begin{frontmatter}

\title{Identification of Parameters for Large-scale Kinetic Models}

\author{Ugur G. Abdulla \& Roby Poteau}
\address{Department of Mathematical Sciences, Florida Institute of Technology, 150 W University Blvd, Melbourne, FL 32901}
\begin{abstract}
Numerical method for the inverse problem on the identification of parameters
for large-scale systems of nonlinear ordinary differential equations (ODEs) arising in systems biology
is introduced. 
The method combines Pontryagin optimization or Bellman's quasilinearization with sensitivity analysis and Tikhonov regularization. 
Embedding a method of staggered corrector for sensitivity analysis and by enhancing multi-objective optimization enables application of the method to large-scale models with practically non-identifiable
parameters based on multiple data sets, possibly with partial and noisy measurements. The method is tested in two canonical 
benchmark models, such as three-step pathway modeled by 8 nonlinear ODEs with 36 unknown and two control input parameters, and 
a model of central carbon metabolism of {\it Escherichia coli} described by a system of 18 linear ODEs with 116 unknown parameters. 
The numerical
results demonstrate superlinear convergence with a minimum data sets and with minimum measurements per data set, and possibly with partial and noisy measurements. Software package {\it qlopt} is developed and posted in GitHub. MATLAB package AMIGO2 is used to demonstrate advantage of {\it qlopt} over most popular methods/software such as {\it lsqnonlin}, {\it fmincon} and {\it nl2sol}.
\end{abstract}


\end{frontmatter}
}]

\section{Introduction}

Systems Biology is an actively emerging interdisciplinary field between biology and mathematics, based on the idea of treating biological systems as a whole entity which is more than the sum of its interrelated components. These systems are networks with emerging properties generated by complex interaction of a large number of cells or organisms.  The mission is to reveal and understand the global properties of biological or bioengineering systems that occur through complex interactions on a microscopic level. One of the major goals of systems biology is to reveal, understand, and predict such properties through the development of mathematical models based on experimental data. In many cases, predictive models of systems biology are described by large systems of nonlinear differential equations. Quantitative identification of such systems requires the solution of inverse problems on the identification of parameters of the system. Inverse problems are ill-posed, meaning that a solution may not exist, may not be unique, and most importantly, may not be continuously dependent on the measurements. Since measurements always contain some error, it is impossible to solve inverse problems without regularization techniques~\cite{Tikhonov, gabor2015robust}.

Ill-posedness of the inverse problems in large scale models of systems biology is strongly associated with the problem of correlation of parameters, expressed in explicit or implicit functional relation between the group of unknown parameters. Correlation of parameters in realistic large scale models is unavoidable, for it is motivated by biological phenomena 
of correlation between dynamics and signatures of genes and proteins. This creates an essential obstacle in solving the inverse problems for many Systems Biology models~\cite{audoly2001global,identif2,identif3,identif4}. The major difficulty arises due to the fact that correlated parameters in general cannot be identified uniquely via measurements. Therefore, correlated parameters are called \textit{non-identifiable}. Besides unknown parameters, large-scale models include control input parameters, which are used for experimental design and measurements. Non-identifiable parameters are called \textit{structurally non-identifiable}, if correlation of parameters is independent of control input parameters. \textit{Structural non-identifiability} of parameters is an intrinsic property of the model and cannot be resolved with additional or more accurate measurements.
On the other hand, if correlation of parameters depends on control input parameters, it is possible that non-identifiable parameter set can be remedied to be identifiable with improved and/or additional measurements. In this case parameters are called \textit{practically non-identifiable}
\cite{Tikhonov,audoly2001global,identif2,identif3,identif4,identif5,identif6}.
Besides the difficulties associated with ill-posedness of  inverse problems,
nonlinearity of the system causes extreme sensitivity with respect to the parameters. Therefore, delicate sensitivity analysis is of high demand \cite{sens}. Inverse problems on the identification of parameters in systems biology are an actively growing research area  \cite{hanke1997regularizing,engl2009inverse,voit2010parameter,voit2,bistable,Yanping_Robust,gentogg,zak,repressilator,amigo2,nl2sol,gabor2015robust,frohlich2017scalable,andrianantoandro2006synthetic}.
We refer to survey articles~\cite{engl2009inverse,raue2013lessons,amigo2,survey2} for the extensive list of references.
Standard popular approach to parameter identification problem in systems biology is its formulation as nonlinear optimization problem with objective to find unknown parameters via minimization of the mismatch or residual between experimental data and model dynamics. Ultimate goal is to develop a global optimization method with least computational cost, which is robust with respect to nonlinearities and scales well with problem size \cite{survey2}. Currently such an ideal method does not exist. Local methods can be classified as gradient-based methods (\cite{schit2013})
and gradient-free ones \cite{wright1996,Conn2009}. Gradient based methods are very effective, but they only provide convergence to local optima. Gradient-free methods are less efficient, and has slow convergence rate \cite{Conn2009}. Global optimization methods can be classified as stochastic (\cite{zhigljavsky2007}) and deterministic ones \cite{esposito2000}. Stochastic global optimization algorithms are implementing pseudorandom sequences to determine search directions toward the global optimum. Most effective class of stochastic methods are pure random search and adaptive sequential methods, clustering methods, population-based methods and nature inspired methods \cite{dreo2006,amigo2}. Advantage of popular global stochastic methods such as simulated annealing, particle swarm optimization, or genetic algorithms is their scalability, but the downside is high computational cost \cite{moles2003parameter}. Deterministic local optimization methods can be used as global optimization method by embedding "Multi-start" strategy into it which facilitates many optimization runs from randomly selected initial parameter guesses \cite{amigo2,raue2013lessons}. Latin hypercube sampling \cite{owen1992} for partition of the parameter space is used to guarantee that each parameter estimation iteration starts with initial guess from different region in parameter space.

Some of the most popular local optimization methods available as
open source software (\cite{amigo2,data2dynamics,pesto}) are 
\begin{itemize}
\item Levenberg-Marquardt algorithm and trust-region-reflective method (function \textit{lsqnonlin} in MatLab) \cite{hanke1997regularizing}; 
\item Sequential Quadratic Programming (function \textit{fmincon} in MatLab Optimization toolbox); 
\item An adaptive non-linear least-squares algorithm (function \textit{nl2sol} in MatLab) \cite{nl2sol}.
\end{itemize}
One can also consider so called {\it hybrid optimization algorithms} as a combination of stochastic and deterministic algorithms: first candidate set of parameter values are generated by stochastic algorithm, followed by deterministic iterative algorithms by choosing candidate set elements as initial guesses \cite{raue2013lessons,amigo2}. In \cite{raue2013lessons}
comprehensive comparison of 15 optimization algorithms from three groups is pursued. 12 stochastic optimization algorithms (\cite{12stochastic}) were compared with a deterministic
trust region algorithm (\cite{coleman96}) in combination with Latin hypecube sampling and two different approaches for calculating derivatives of the objective function: finite difference approximation and analytically derived sensitivity equations.  The results show that multi-start deterministic optimization using the sensitivity equations for the calculation of derivatives significantly outperforms all other tested algorithms. The performance of stochastic optimization algorithms was surprisingly low compared to the hybrid and fully deterministic optimization algorithms \cite{raue2013lessons} .

In another recent paper \cite{survey2} collection of benchmark problems were selected to evaluate the performance of two families of optimization methods:
\begin{itemize}
\item  multi-start local optimization (MS),
\item enhanced scatter search metaheuristic (eSS),
\end{itemize}
the latter may be combined with deterministic local searches, leading to hybrid methods. Selected local optimization methods were
\begin{itemize}
\item {\it nl2sol}-FWD: the nonlinear least-squares algorithm {\it nl2sol}, using forward sensitivity analysis for evaluating the gradients of the residuals.
\item {\it fmincon}-ADJ: the interior point algorithm included in {\it fmincon}, using adjoint sensitivities for evaluating the gradient of the objective function.
\item {\it dhc}-a gradient-free dynamic hill climbing algorithm.
\item eSS without any local methods (NOLOC) and particle swarm optimization (PSO) \cite{kennedy1995}.
\end{itemize}
Comprehensive evaluation of \cite{survey2} clearly shows that high-quality sensitivity calculation methods provide a competitive advantage to local methods that exploit them. Optimization using adjoint and forward sensitivity analysis ${\it fmincon}-ADJ$ and ${\it nl2sol}-FWD$ usually outperform the gradient-free alternative {\it dhc}.
 The combinations of eSS with gradient-based methods, eSS-{\it fmincon}-ADJ and eSS-{\it nl2sol}-FWD, clearly outperform the gradient-free alternatives, eSS-{\it dhc} and eSS-NOLOC, as well as the gradient-free PSO.

 The comparison analysis performed in \cite{raue2013lessons,survey2} demonstrates that robust deterministic local optimization methods embedded with MS or eSS strategy, and with sharp sensitivity analysis platform are the best candidates for creation of powerful global optimization methods for large-scale models of systems biology.

The goal of this paper is to develop effective local optimization method for solving inverse problems on the identification of finite-dimensional parameter sets for the large-scale systems of nonlinear ODEs arising in Systems Biology, and to develop an effective software which is competitive with currently known popular methods/software such as {\it lsqnonlin}, {\it fmincon} and {\it nl2sol}.

In \cite{previouspaper} we implemented the numerical method, introduced originally in~\cite{abdulla1,abdulla2}, for the solution of the inverse problems for the canonical models of Systems Biology.
The iterative method combines ideas of Pontryagin's optimization or Bellman's quasilinearization with sensitivity analysis and Tikhonov regularization. Extensive computational analysis pursued in~\cite{previouspaper}  demonstrates that the method is very well adapted to canonical models of system biology with moderate size parameter sets and has quadratic convergence. Software package \textit{qlopt} was developed and posted in GitHub~\cite{qlopt}. MATLAB package AMIGO2~\cite{amigo2} was used to demonstrate the competitiveness and advantage of \textit{qlopt} with other most popular local search methods like \textit{lsqnonlin, fmincon, nl2sol}.

However, direct adaptation and scalability of the method implemented in \cite{previouspaper}  to inverse problems with significantly larger size was not as effective. In this paper we introduce a modification of the method which is effective to solve the inverse problem on the identification of parameters for large scale models in systems biology. In Section~\ref{method} we introduce the new modified method accompanied with two types of Tikhonov regularization algorithms. The modification is twofold.
\begin{itemize}
\item Method of staggered corrector \cite{staggeredCorrector} is embedded into the step of calculation of the sensitivity vectors. Precisely, instead of solving linearized system and associated sensitivity system, we first solve original system through quasilinearization \cite{bellman}, and then use its solution to solve linear sensitivity system corresponding to the original nonlinear system. We use software package CVODES \cite{CVODES} to implement the method of staggered corrector into our algorithm.
\item Multi-objective optimization is added to the method which enables the application of the method to large-scale models that are practically non-identifiable due to parameter correlation.
\end{itemize}
In Section~\ref{Results and Discussions} we present results and analysis of the application of the method to benchmark kinetic model of a biological network for a three-step pathway modeled by 8 nonlinear ODEs describing 8 metabolic concentrations with 36 unknown parameters and two control input parameters specifying the experimental design. Section~\ref{Numerical Results with Noise-free Data Sets} presents the results for noise-free simulated data. It is demonstrated that 5 to 16 data sets are satisfactory to uniquely identify all 36 parameters with high precision.
This is followed by Section~\ref{Effect of the Regularization Parameter} where we demonstrate that the delicate implementation of the Tikhonov regularization with optimal choice of the regularization parameter significantly affects the convergence rate and precision of the algorithm. In the following Section~\ref{Convergence vs. Number of Data Points} we demonstrate the effect of the number of time measurements for each component of the system on the convergence and accuracy of the method. Subsequently in Section~\ref{Convergence vs. Number of Data Sets} we explore in depth the major question of identifiability of the parameters with minimum number of data sets. It is demonstrated that minimum five data sets are required to identify uniquely all parameters with high precision. Section~\ref{Range of convergence} demonstrates that the careful implementation of the Type II Tikhonov regularization significantly improves the convergence range of the algorithm.
In Section~\ref{Convergence with Noisy Measurements} we apply the method to simulated noisy data  and demonstrate its robustness. Section~\ref{cr} presents numerical analysis of the rate of convergence of the method. Section~\ref{Convergence with Partial Measurements} demonstrates the robustness of the method with respect to partial measurements.
In Section~\ref{comparison}, we pursue comparison and demonstrate the competitiveness and the advantages of our method and the associated software package \textit{qlopt} against most advance methods/software like \textit{lsqnonlin, fmincon, nl2sol}. 

In Section~\ref{B2} we analyze benchmark kinetic model of central carbon metabolism of {\it Escherichia coli} 
modeled by 18 linear differential equations for the concentrations of 17 intracellular metabolites and extracellular glycose \cite{b2model,benchmarkpaper}. We apply the method to inverse problem on the identification of 116 parameters which express kinetic properties and maximum reaction rates. Section~\ref{b2nonoise} presents the results with noise-free simulated data. High accuracy and robustness of the method with respect to number of time-measurements both in linear and logarithmic scale, convergence with partial measurements, and superlinear convergence rate is demonstrated. In Section~\ref{b2noise} we apply the method to simulated noisy data  and demonstrate its robustness. Section~\ref{b2rc} demonstrates that the careful implementation of the Type II Tikhonov regularization significantly increases convergence range of the method in logarithmic scale.

Finally, in Section~\ref{conclusions} we outline the main conclusions.

\section{Description of the Method}\label{method}
Consider a dynamical system:
\begin{equation} \label{eq:one}
	\frac{\mathrm d \bf{x}}{\mathrm d t} = \mathbf{f}(t, \mathbf{x}, \mathbf{u}, \mathbf{v}), \ t_0\leq t \leq t_1
\end{equation}
\begin{equation} \label{eq:two}
	\mathbf{x}(t_0) = \mathbf{x}^0 \in \mathbb{R}^n,
\end{equation}
where
\[ \mathbf{x} = \mathbf{x}(t)= (x_1(t), x_2(t), \ldots, x_n(t)): [t_0, t_1] \rightarrow \mathbb{R}^n\]
is the state vector,
$$\mathbf{u} = (u_1, u_2, \ldots, u_m)  \in \mathbb{R}^m$$
is the unknown parameter vector,
$$\mathbf{v} = (v_1, v_2, \ldots, v_p)  \in \mathbb{R}^p$$
is the control input parameter vector, and
\begin{align*}
\mathbf{f}=(f_1&(t,  \mathbf{x},  \mathbf{u}, \mathbf{v}), f_2(t,  \mathbf{x},  \mathbf{u}, \mathbf{v}), \ldots, f_n(t,  \mathbf{x},  \mathbf{u}, \mathbf{v})):\\
			&[t_0, t_1] \times \mathbb{R}^n \times \mathbb{R}^m \times \mathbb{R}^p \rightarrow \mathbb{R}^n
\end{align*}
is a continuous vector function with continuous derivatives
\[ \frac{\partial \mathbf{f}}{\partial \mathbf{x}}, \  \frac{\partial \mathbf{f}}{\partial \mathbf{u}}. \]

Consider {\it inverse problem of finding the parameter $\mathbf{u}$ given $D$ measurements for the state vector $\mathbf{x}$ corresponding to $D$
fixed values of the control vector $\mathbf{v}$:
$$\mathbf{x} = \mathbf{x}^d(t)= \mathbf{x}^d(t; \mathbf{u}) := \mathbf{x}(t,\mathbf{u}, \mathbf{v}^d), \ d=1,...,D$$
on an interval $t_0 \leq t \leq t_1$, where $ \mathbf{x}^d(t_0) =  \mathbf{x}^0$.}

Having chosen the initial vector function $\mathbf{x}^d_{N,0}$ (say, $\mathbf{x}^d_{N,0}=\mathbf{x}^d(t)$), and initial approximation $\mathbf{u}=\mathbf{u}_0$, we implement quasilinearization of \eqref{eq:one} (\cite{bellman}) and at each fixed iteration $N=1,2,...$ we find the solution as a limit
\begin{equation}\label{newton}
\mathbf{x}^d_{N}(t) = \lim_{p\to \infty}\mathbf{x}^d_{N,p}(t), \ t_0\leq t \leq t_1,
\end{equation}
where $\mathbf{x}^d_{N,p}$ solves the linear system of ODEs in $[t_0,t_1]$ with $\mathbf{u}=\mathbf{u}_{N-1}$:
\begin{subequations}\label{quasi}
	\begin{gather}
	\frac{\mathrm d \mathbf{x}^d_{N,p}}{\mathrm d t} = \mathbf{f}(t, \mathbf{x}^d_{N,p-1}, \mathbf{u}, \mathbf{v}^d)+\nonumber\\
	J(t, \mathbf{x}^d_{N,p-1}, \mathbf{u}, \mathbf{v}^d)(\mathbf{x}^d_{N,p} - \mathbf{x}^d_{N,p-1}), \label{quasi1}\\
	\mathbf{x}^d_{N,p}(t_0) = \mathbf{x}^0,\label{quasi2}
	\end{gather}
\end{subequations}
where
\[ J(t, \mathbf{x}, \mathbf{u}, \mathbf{v}) = \frac{\partial \mathbf{f}(t, \mathbf{x}, \mathbf{u}, \mathbf{v})}{\partial \mathbf{x}}\]
is the n$\times$n Jacobian matrix. It is well known that the convergence \eqref{newton} has a quadratic rate \cite{bellman}.
Given the initial guess $\mathbf{u}_0$ of the unknown parameter $\mathbf{u}$, we identify at every step of the iteration a new approximation
\begin{equation}\label{update}
\mathbf{u}_N=\mathbf{u}_{N-1}+\Delta \mathbf{u},
\end{equation}
which minimizes the $L_2$-norm of the residues
$${\cal R}=\mathbf{x}^d(t, \mathbf{u}) - \mathbf{x}^d_{N}(t, \mathbf{u}_{_{N}}).$$
We have
\begin{equation}\label{eq:sens}
	{\cal R} = \Delta \mathbf{x}^d_N(t) -  U^d_N \Delta \mathbf{u} +o(|\Delta \mathbf{u}|), \quad\text{as}~|\Delta \mathbf{u}|\to 0,
\end{equation}
where
\[ \Delta \mathbf{x}^d_{N}(t) = \mathbf{x}^d(t, \mathbf{u}) - \mathbf{x}^d_N(t, \mathbf{u}_{N-1}), \]
$U^d_N$ is an $n\times m$ sensitivity matrix with columns
\[ U_N^{d,j}=\Big ( \frac{\partial \mathbf{x}^d_N(t,\mathbf{u}_{N-1})}{\partial u^j}\Big ), \ j=1,...,m. \]
$U_N^d$ solves the matrix differential system

\begin{subequations} \label{eq:senssys}
	\begin{gather}
	\frac{dU^d_N}{dt} = \frac{\partial}{\partial \mathbf{u}} \mathbf{f}(t, \mathbf{x}^d_{N}, \mathbf{u}_{N-1}, \mathbf{v}^d) + \nonumber\\
	J(t, \mathbf{x}^d_{N}, \mathbf{u}_{N-1}, \mathbf{v}^d)U^d_N, \ t_0\leq t \leq t_1  \label{eq:unsys}\\
	U^d_N(t_0) = 0,\label{eq:uninit}
	\end{gather}
\end{subequations}
where \(\mathbf{x}^d_N\) is the solution of \eqref{eq:one},\eqref{eq:two} with $\mathbf{u}=\mathbf{u}_{N-1}, \mathbf{v}=\mathbf{v}^d$
as it is constructed in \eqref{newton}. Finding $\mathbf{x}^d_N, U_N^d$ from \eqref{newton}, \eqref{quasi}, \eqref{eq:senssys} form the method of staggered corrector \cite{staggeredCorrector}.

To find $\Delta \mathbf{u}$, we minimize the multi-objective function
\begin{equation} \label{eq:multiobjective}
	\mathcal{J}(\Delta \mathbf{u}) = \sum_{d=1}^{D}||\Delta \mathbf{x}^d_{N} - U^d_{N} \Delta \mathbf{u}||^{2}_{L_2^n(t_0,t_1)},
\end{equation}
where $L^n_2(t_0,t_1)$ is a Hilbert space of vector functions $g:(t_0,t_1) \to \mathbb{R}^n$ with inner product
\[ (g,h)_{L^n_2(t_0,t_1)} =\int_{t_0}^{t_1}g^Thdt. \]
We have
\[ \mathcal{J}_N'(\Delta \mathbf{u}) =  2\sum_{d=1}^{D}\int\limits_{t_0}^{t_1} \Big [ (U^d_N)^T U^d_N \Delta \mathbf{u} - (U^d_N)^T \Delta \mathbf{x}^d_{N}\Big ]  \mathrm{d}t,\]
\[ \mathcal{J}_N''(\Delta \mathbf{u}) = 2\sum_{d=1}^{D}\int\limits_{t_0}^{t_1} (U^d_N)^T U^d_N\mathrm{d}t.\]
Therefore, minimum $\Delta\mathbf{u}$ satisfies the following system of linear algebraic equations
\begin{equation}\label{eq:nece}
	A_N\Delta \mathbf{u} = P_N,
\end{equation}
where
\[ A_N =  \sum_{d=1}^{D}\int\limits_{t_0}^{t_1} (U^d_N)^T U^d_N \mathrm{d}t=\Big (a_N^{ij}\Big )_{i,j=1}^m   \]
is an $m\times m$ symmetric matrix with elements
\[ a_N^{ij}=\sum_{d=1}^{D}\int_{t_0}^{t_1}\Big (\frac{\partial x_N^d(t,\mathbf{u}_{N-1})}{\partial u^i}\Big )^T \frac{\partial x_N^d(t,\mathbf{u}_{N-1})}{\partial u^j}dt, \]
and
\[  P_N =  \sum_{d=1}^{D}\int\limits_{t_0}^{t_1} (U^d_N)^T \Delta \mathbf{x}_{N} \mathrm{d}t=\Big (p_N^j\Big )_{j=1}^m \]
is an $m$-vector with elements
\begin{gather*} p_N^j=\sum_{d=1}^{D}\int_{t_0}^{t_1}\Big (\frac{\partial x_N^d(t,\mathbf{u}_{N-1})}{\partial u^j}\Big )^T(x^d(t,\mathbf{u})\\
-x_N^d(t,\mathbf{u}_{N-1}))dt.
\end{gather*}
In fact, $A_N$  is a sum of Gram matrices $A_N^d$ of vectors $U_N^{d,j}$, and
\[ a_N^{ij}=\sum_{d=1}^{D}(U_N^{d,i},U_N^{d,j})_{L^n_2(t_0,t_1)}. \]
It is known \cite{beckenbach} that
\[ det (A_N^d)=\Gamma (U^{d,1}_N,...,U^{d,m}_N)\geq 0 \]
and it is positive, that is to say, $A_N^d$ is non-singular, if and only if the vectors $U_N^{d,j}, j=1,...,m$ are linearly independent.

Hence, we suggest the following modfication of the numerical algorithm from~\cite{abdulla1}.
\subsection{Algorithm}\label{algorithm}
\begin{enumerate}
  	\item Initialize $\mathbf{u}_0$ and set $N=1$.
  	\item Set $\mathbf{x}^d_{N,0}(t)$  and find $\mathbf{x}^d_N(t, \mathbf{u}_{N-1})$ via quasilinearization from~\eqref{newton},\eqref{quasi}.
	\item Having $\mathbf{x}^d_N(\cdot, \mathbf{u}_{N-1})$ find sensitivity matrices $U^d_N$ by solving linear ODE system~\eqref{eq:senssys}.
 	\item Find $\Delta \mathbf{u}$ by solving linear algebraic equations system~\eqref{eq:nece} and update the new value $\mathbf{u}_N$ of the parameter using~\eqref{update}.
	\item If satisfactory accuracy is achieved, then terminate the process, otherwise replace $N$ with $N+1$ and go back to Step 2. As termination criteria, the smallness of either of the expressions
	\[ |\mathbf{u}_{_{N-1}}-\mathbf{u}_{_N} |, \ \mathcal{J}_N(\Delta \mathbf{u}), \ \sum_{d=1}^{D} ||\mathbf{x}^d(\cdot)-\mathbf{x}^d_N(\cdot,\mathbf{u}_N)||_{L^n_2}\]
can be used.
	\end{enumerate}

\subsection{Regularization}\label{regularization}
As in \cite{previouspaper} we implement two types of Tikhonov regularization. Type I regularization is performed by replacing the function
\eqref{eq:multiobjective} with
\begin{equation} \label{eq:regone}
	\sum_{d=1}^{D}\left(||\Delta \mathbf{x}^d_{N} -U^d_N \Delta \mathbf{u}||^2_{L_2^n}\right) +\alpha | \Delta \mathbf{u}|^2.
\end{equation}
This yields the following linear system instead of \eqref{eq:nece}
\begin{equation} \label{eq:regtype1}
	(A_N +\alpha I)\Delta \mathbf{u} = P_N
\end{equation}
where $I$ is the identity matrix and $\alpha$ is a regularization parameter.
Type II regularization is performed by replacing the function \eqref{eq:multiobjective} with
\begin{equation} \label{eq:regtwo}
		\sum_{d=1}^{D} ||\Delta\mathbf{x}^d_{N} -U^d_N \Delta \mathbf{u}||^2_{L_2^n} +\alpha | \mathbf{u}_{N-1} + \Delta \mathbf{u}  - \mathbf{u}^{*}|^2
\end{equation}
where $\mathbf{u}^{*}$ is a known vector expected to be close to the true value of the unknown parameter. This implies the following linear system instead of \eqref{eq:nece}:
\begin{equation} \label{eq:regtype2}
	(A_N + \alpha I) \Delta \mathbf{u} = P_N + \alpha(\mathbf{u}^{*} - \mathbf{u}_{N-1}).
\end{equation}

\subsection{Identifiability vs. Practical Non-identifiabilty}\label{identifiability}
Convergence of the algorithm is connected to the identifiability of unknown parameters. In fact, $d$th Gram matrix summand $A_N^d$ of $A_N$ in \eqref{eq:nece} is
called Fisher information matrix (FIM) for the ODE system \eqref{eq:one}, which characterizes the information content of the experimental measurement in the $d$th data set.
Singularity of $A_N^d$ is equivalent to linear dependence of the sensitivity vectors $U_N^{d,j}, j=1,...,m$. From \eqref{eq:senssys} it follows that the latter is equivalent to linear dependence of the columns of the matrix $\frac{\partial}{\partial \mathbf{u}} \mathbf{f}(t, \mathbf{x}^d_{N}, \mathbf{u}_{N-1}, \mathbf{v}^d)$, which is, in general, equivalent to correlation of parameters. Hence, singularity of the matrix $A_N$ is equivalent to non-identifiability of parameters. However, note that the matrix $\frac{\partial \mathbf{f}}{\partial \mathbf{u}}$ depends on the control input parameter $\mathbf{v}$. Therefore, if selection of ${\bf v}$ can arrange independence of the columns of the matrix $\frac{\partial \mathbf{f}}{\partial \mathbf{u}}$, the correlation of parameters will be remedied, and parameters will be {\it practically non-identifiable}. This will guarantee non-singularity of $A_N$, and convergence of the algorithm can be established. The major goal is to pursue experimental design and collection of measurements for carefully selected values of the control parameter $\mathbf{v}$, so that the algorithm converges to the unique solution of the inverse problem. Yet another difficulty arises when $A_N$ is non-singular but $det A_N$ is sufficiently small, and for computer simulation $A_N$ is treated as a singular matrix \cite{ljung,zak,identif6}. Our two regularization algorithms are developed to address such {\it practical non-identifiability} cases.
A major factor for the convergence of the algorithm for the identification of {\it practically non-identifiable} parameters is the increase of number of data sets $D$. Specifically, there is a minimum number of data sets with different inputs of control parameters for experimental design needed to relieve the parameter correlations and acquire suitable measurement data for unique parameter estimation \cite{identif2}.


\section{Benchmark Model 1: Biological Network for a Three-step Pathway}\label{Results and Discussions}
We tested the method on a benchmark model of a biological network for a three-step pathway modeled
by 8 nonlinear ODEs describing 8 metabolic concentrations and 36 parameters $p_i, i=1,...,36$ (\cite{mendes20018}).
Two parameters $P$ and $S$ are control input parameters specified by the experimental design.
The unknown parameters $p_i$ are correlated, but their functional relationship with one another
is dependent on the input parameters \(P\) and \(S\), and in general parameters are {\it practically non-identifiable}, and can be identified
with multiple data sets. In \cite{moles2003parameter}, the inverse problem was analyzed with 16 noise-free data sets, and in \cite{identif6}
with 16 both noise-free and noisy data sets. The results demonstrated strong parameter correlations in several groups, with accurate parameter values identified in \cite{identif6}. Parameter correlations were analyzed in \cite{identif2}.
It is demonstrated that correlated parameters are practically non-identifiable for a single data set and at least 5 data sets
with different control inputs are required to uniquely estimate the 36 parameters of this model.

\begin{align*}
	\dot{x}_1 & = \frac{p_1}{1 + {\left(\frac{P}{p_2}\right)}^{{p}_{3}} + {\left(\frac{p_{4}}{S}\right)}^{p_5}} - p_6x_1 \\
	\dot{x}_2 & = \frac{p_7}{1 + {\left(\frac{P}{p_8}\right)}^{p_{9}} + {\left(\frac{p_{10}}{x_7}\right)}^{p_{11}}} - p_{12}x_2 \\
	\dot{x}_3 & = \frac{p_{13}}{1 + {\left(\frac{P}{p_{14}}\right)}^{p_{15}} + {\left(\frac{p_{16}}{x_8}\right)}^{p_{17}}} - p_{18}x_3 \\
	\dot{x}_4 & = \frac{p_{19}x_1}{p_{20} + x_1} -  p_{21}x_4 \\
	\dot{x}_5 & = \frac{p_{22}x_2}{p_{23} + x_2} -  p_{24}x_5 \\
	\dot{x}_6 & = \frac{p_{25}x_3}{p_{26} + x_3} -  p_{27}x_6 \\
	\dot{x}_7 & = \frac{p_{28}x_4 \left(S - x_7 \right)}{p_{29}\left(1 + \frac{S}{p_{29}} + \frac{x_7}{p_{30}}\right) } - \frac{p_{31}x_5 \left(x_7 - x_8 \right)}{p_{32}\left(1 + \frac{x_7}{p_{32}} + \frac{x_8}{p_{33}}\right) } \\
	\dot{x}_8 & = \frac{p_{31}x_5 \left(x_7 - x_8 \right)}{p_{32}\left(1 + \frac{x_7}{p_{32}} + \frac{x_8}{p_{33}}\right) } - \frac{p_{34}x_6 \left(x_8 - P \right)}{p_{35}\left(1 + \frac{x_8}{p_{35}} + \frac{P}{p_{36}}\right) }
\end{align*}

For our experiments we used the common values for the initial conditions
\((6.6667e-1, 5.7254e-1, 4.1758e-1, 4.0e-1,3.6409e-1, 2.9457e-1, 1.419,
9.3464e-1)\), with \(t_0 = 0\) and \(t_1 = 120.\) True values of 36 parameters are outlined in last columns of Tables~\ref{tab:tsmn_num_of_data_sets} and   ~\ref{tab:tsmn_21_num_of_data_sets} in Appendix. We implemented 16 input parameters given in AMIGO2 \cite{amigo2} (Table~\ref{tab:control_params_16} in Appendix) and 5 input parameters given in \cite{identif2} (Table~\ref{tab:control_params_05} in Appendix) for our experiments. We chose the regularization parameter $\alpha$ as a function of the residual:
\begin{equation}\label{eq:alphaformula}
	\sum_{d=1}^{D} C ||\mathbf{x}^{d} - \mathbf{x}^d_{N} ||^{\gamma}_{L_2}
\end{equation}
where \(C,\gamma >0\) are chosen experimentally.

\subsection{Numerical Results with Noise-free Data Sets}\label{Numerical Results with Noise-free Data Sets}
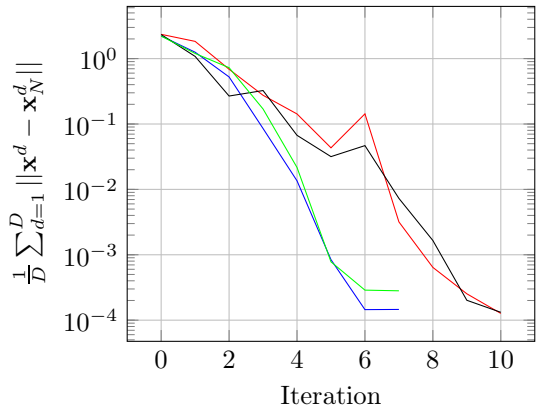
\begin{figure}
\begin{tikzpicture}
\begin{semilogyaxis}[
	xlabel=Iteration,
	ylabel=\(\frac{1}{D}\sum^D_{d=1}||\mathbf{x}^{d} - \mathbf{x}^{d}_{N}||\),
	grid=major,
	legend style={font=\tiny}
]
\addplot[color=red,grid=major] coordinates {
(0,2.3531547)
(1,1.83531849)
(2,0.68798669)
(3,0.274885327)
(4,0.142481187)
(5,0.0432371423)
(6,0.142963015)
(7,0.00318242338)
(8,0.000640835211)
(9,0.000253128874)
(10,0.000126816252)
}; 
\addplot[color=blue,grid=major] coordinates {
(0,2.22189181)
(1,1.25344256)
(2,0.52453039)
(3,0.0853258063)
(4,0.0135211642)
(5,0.000842377264)
(6,0.000144916284)
(7,0.000145922839)
}; 
\addplot[color=black,grid=major] coordinates {
(0,2.33449974)
(1,1.08054098)
(2,0.267827022)
(3,0.32485918)
(4,0.0671332181)
(5,0.0317082122)
(6,0.0467198593)
(7,0.0072885792)
(8,0.00165235743)
(9,0.000201427169)
(10,0.000131889726)
}; 
\addplot[color=green,grid=major] coordinates {
(0,2.21372838)
(1,1.19907178)
(2,0.727159858)
(3,0.169737496)
(4,0.0216480157)
(5,0.00078048984)
(6,0.000287942554)
(7,0.000281180064)
}; 
\end{semilogyaxis}
\end{tikzpicture}
\caption{The average error at each iteration with
(green) 16 data sets, \(t_0 = 0\), \(t_1 = 120\), \(\Delta t = 0.5\), i.e. 240 time points. Regularization parameter \(\alpha\) was determined using \eqref{eq:alphaformula} where \(C = 0.009\) and \(\gamma = 2\);
(black) with 5 data sets, \(t_0 = 0\), \(t_1 = 120\), \(\Delta t = 0.5\), i.e. 240 time points. Regularization parameter \(\alpha\) was determined using \eqref{eq:alphaformula} where \(C = 0.25\) and \(\gamma = 1\);
(blue) with 16 data sets, \(t_0 = 0\), \(t_1 = 120\), \(\Delta t = 6.0\), i.e. 20 time points. Regularization parameter \(\alpha\) was determined using \eqref{eq:alphaformula} where \(C = 0.005\) and \(\gamma = 2\);
(red) with 5 data sets, \(t_0 = 0\), \(t_1 = 120\), \(\Delta t = 6.0\), i.e. 20 time points. Regularization parameter \(\alpha\) was determined using \eqref{eq:alphaformula} where \(C = 0.25\) and \(\gamma = 2\).}
\label{fig:tsmn_no_noise_combo}
\end{figure}

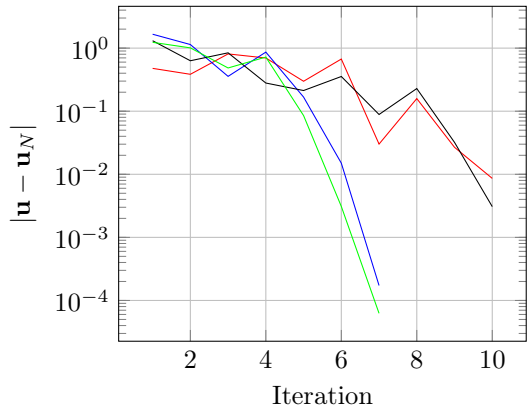
\begin{figure}
\begin{tikzpicture}
\begin{semilogyaxis}[
	xlabel=Iteration,
	ylabel=\(|\mathbf{u} - \mathbf{u}_{N}|\),
	grid=major,
	legend style={font=\tiny}
]
\addplot[color=red,grid=major] coordinates {
(1,0.477401898)
(2,0.384626757)
(3,0.809039599)
(4,0.697700566)
(5,0.298342159)
(6,0.671089205)
(7,0.0301312763)
(8,0.15848881)
(9,0.0267651592)
(10,0.00860838691)
}; 
\addplot[color=blue,grid=major] coordinates {
(1,1.66071082)
(2,1.13964911)
(3,0.356689066)
(4,0.864109038)
(5,0.167496018)
(6,0.0150177003)
(7,0.000172676348)
}; 
\addplot[color=black,grid=major] coordinates {
(1,1.30966588)
(2,0.630654924)
(3,0.842198026)
(4,0.27963278)
(5,0.212711287)
(6,0.354916998)
(7,0.088721767)
(8,0.228530229)
(9,0.0320505864)
(10,0.00307372189)
}; 
\addplot[color=green,grid=major] coordinates {
(1,1.2352208)
(2,1.00873259)
(3,0.484530274)
(4,0.721368658)
(5,0.085288664)
(6,0.00309611878)
(7,6.25232925e-05)
}; 
\end{semilogyaxis}
\end{tikzpicture}
\caption{The average parameter error at each iteration with the same configurations as in Figure~\ref{fig:tsmn_no_noise_combo}.}
\label{fig:tsmn_no_noise_combo_params}
\end{figure}

We applied the numerical method to identify the 36 parameters with 16 and 5 data sets. We generated simulated measurements for each data set by solving the system of 8 nonlinear ODEs with true values of 36 parameters. We chose the number of time data points for each of the 8 components of the system either at 240 or at 20 uniformly distributed time grid points in the segment $[0, 120]$. Computational cost of each iteration per one data set consists of iterative solution of the system of 8 nonlinear ODEs through quasilinearization; solving a system of 288 linear ODEs to identify sensitivity matrix-function; calculation of 1332 integrals for entries of the matrix $A_N$ and vector $P_N$; and finally solving a  system of 36 linear algebraic equations to find the increment of the parameters. The green line in Figures~\ref{fig:tsmn_no_noise_combo} and~\ref{fig:tsmn_no_noise_combo_params} demonstrate the results for 16 data sets with 240 time points. Rapid convergence to the true solution happens in only 7 iterations. Next we applied the method with 5 data sets. Though it required 3 extra iterations, the black line in Figures~\ref{fig:tsmn_no_noise_combo} and~\ref{fig:tsmn_no_noise_combo_params} demonstrate the rapid convergence of the method with reduced error.

Next we applied the method by choosing measurements at 20 time grid points for each of the 8 components. The results are demonstrated for 16 and 5 data sets in Figures~\ref{fig:tsmn_no_noise_combo} and~\ref{fig:tsmn_no_noise_combo_params} by the blue and red lines, respectively. The algorithm converges in the same number of iterations with respect to the number of data sets, while maintaining around the same level of accuracy, as demonstrated in Figures~\ref{fig:tsmn_no_noise_combo} and~\ref{fig:tsmn_no_noise_combo_params}.

\subsection{Effect of the Regularization Parameter \(\alpha\)}\label{Effect of the Regularization Parameter}
The choice of the regularization parameter $\alpha$ is an important factor which significantly improves the convergence rate
and computational cost of the algorithm. To demonstrate the existence of the optimal non-trivial value of $\alpha$ at every fixed step $N$, we considered profiles of \(\alpha\) vs \( |\mathbf{u}_{N-1} + \Delta \mathbf{u} - \mathbf{u}| \), where \(\mathbf{u}\)  is the true solution. Figure~\ref{fig:tmsn01_alpha_all} corresponds to the 2nd and 4th iterations of the green line in Figure~\ref{fig:tsmn_no_noise_combo}. Similarly, Figure~\ref{fig:tmsn02_alpha_all} corresponds to 3rd and 6th iteration of the black line in the same figure. In each example there is a clear minimum which is the best choice of the regularization parameter.
The bullets on the graph corresponds to our choice of the regularization parameter according to the residual method \eqref{eq:alphaformula}. In fact, optimal or nearly optimal choice of the regularization parameter significantly increases convergence rate of the method from geometric to be close to quadratic convergence (see Section~\ref{cr}). The residual method provides a close, but not necessarily optimal value of $\alpha$. Figure~\ref{fig:tmsn01_alpha_all} demonstrates that our choice of the regularization parameter by the residual method is optimal. However, Figure~\ref{fig:tmsn02_alpha_all} demonstrates that residual method doesn't necessarily provide optimal choice of $\alpha$. This analysis demonstrates that there is room for improvement of the convergence rate of the algorithm through implementation of a more effective method for the search of regularization parameter $\alpha$ without significantly affecting computational cost.
\input{./figs/tsmn_no_noise_16ds_alpha_all.tex}
\input{./figs/tsmn_no_noise_05ds_alpha_all.tex}

\subsection{Convergence vs. Number of Data Points}\label{Convergence vs. Number of Data Points}
The method is very robust and convergence is still the case if the number of data points is reduced to a single time measurement at the end of the time interval for each of the
8 components of the system. Figure~\ref{fig:tsmn_data_pnts} demonstrates the dependence of the number of time measurements for each component on the average error
\[ \frac{1}{D}\sum^D_{d=1}||\mathbf{x}^{d} - \mathbf{x}^{d}_{N}||_{L_2^n}\]
calculated at the final iteration in the experiment with $D=5$ data sets. Three graphs correspond to three different settings of the relative and absolute tolerances for CVODES.
Decrease of the latter increases the overall accuracy of the result. Similar dependence in the experiment with 16 data sets and with CVODES tolerance being set up at \(\num{1E-6}\)
is demonstrated in Figure~\ref{fig:tsmn_data_pnts_16}. Some of the variation in the chart can be attributed to error accumulation and noise. 
\begin{figure}
\begin{tikzpicture}
\begin{loglogaxis}[
xlabel=\# of data points per data set,
ylabel=\(\frac{1}{D}\sum^D_{d=1}||\mathbf{x}^{d} - \mathbf{x}^{d}_{N}||\),
grid=major,
legend style={font=\tiny}
]
\addplot[color=blue,grid=major] coordinates {
(1,0.00412806)
(2,0.000135958)
(3,6.42295e-05)
(4,4.397e-05)
(5,2.97971e-05)
(6,3.8912e-05)
(8,3.75759e-05)
(10,2.41929e-05)
(12,4.67753e-05)
(15,3.92436e-05)
(20,3.69709e-05)
(24,3.59108e-05)
(30,3.71826e-05)
(40,3.93489e-05)
(60,3.57601e-05)
(120,3.63562e-05)
(160,3.60404e-05)
(240,3.60359e-05)
(480,3.63819e-05)
};
\addlegendentry{$\num{1E-6}$}
\addplot[color=red,grid=major] coordinates {
(1,8.8604e-06)
(2,2.13553e-06)
(3,3.73645e-06)
(4,5.25056e-06)
(5,5.17534e-06)
(6,5.40193e-06)
(8,4.83586e-06)
(10,5.87531e-06)
(12,5.89593e-06)
(15,4.30799e-06)
(20,6.6774e-06)
(24,4.97908e-06)
(30,7.08137e-06)
(40,7.0175e-06)
(60,6.67377e-06)
(120,7.02305e-06)
(160,5.64319e-06)
(240,6.36273e-06)
(480,6.28476e-06)
};
\addlegendentry{$\num{1E-7}$}
\addplot[color=black,grid=major] coordinates {
(1,3.48837e-07)
(2,3.73059e-07)
(3,8.745e-07)
(4,5.59738e-07)
(5,3.65585e-07)
(6,4.54054e-06)
(8,5.11975e-07)
(10,5.93768e-07)
(12,6.66659e-07)
(15,1.02828e-06)
(24,8.09838e-07)
(20,8.23309e-07)
(30,9.11835e-07)
(40,1.10809e-06)
(60,1.16046e-06)
(120,1.12025e-06)
(160,1.28849e-06)
(240,1.17256e-06)
(480,1.1436e-06)
};
\addlegendentry{\(\num{1E-8}\)}
\end{loglogaxis}
\end{tikzpicture}
\caption{Graph of the number data points per data set vs.
\(\frac{1}{D}\sum^D_{d=1}||\mathbf{x}^{d} - \mathbf{x}^{d}_{N}||\) using 5 data sets, where
$\mathbf{u}_0 = 1.25\mathbf{u}$.}
 Relative and absolute tolerance for CVODES set to \(\num{1E-6}\), \(\num{1E-7}\), and \(\num{1E-8}\).
\label{fig:tsmn_data_pnts}
\end{figure}
\begin{figure}
\begin{tikzpicture}
\begin{loglogaxis}[
xlabel=\# of data points per data set,
ylabel=\(\frac{1}{D}\sum^D_{d=1}||\mathbf{x}^{d} - \mathbf{x}^{d}_{N}||\),
grid=major,
legend style={font=\tiny}
]
\addplot[color=red,grid=major] coordinates {
(1,3.73812e-05)
(2,2.72241e-05)
(3,4.94505e-05)
(4,4.40163e-05)
(5,8.40475e-05)
(6,3.64675e-05)
(8,3.95152e-05)
(10,4.46297e-05)
(12,3.90467e-05)
(15,4.39901e-05)
(24,4.37917e-05)
(30,4.50356e-05)
(40,4.82293e-05)
(60,4.64188e-05)
(120,4.86261e-05)
(160,4.68451e-05)
(240,4.8295e-05)
(440,4.85936e-05)
};
\end{loglogaxis}
\end{tikzpicture}
\caption{Graph of the number time points vs.
\(\frac{1}{D}\sum^D_{d=1}||\mathbf{x}^{d} - \mathbf{x}^{d}_{N}||\) using 16 data sets, , where
$\mathbf{u}_0 = 1.25\mathbf{u}$.} Relative and absolute tolerance for CVODES set to \(\num{1E-6}\)
\label{fig:tsmn_data_pnts_16}
\end{figure}

\subsection{Convergence vs. Number of Data Sets}\label{Convergence vs. Number of Data Sets}
Our numerical analysis confirms the result of \cite{identif2} that at least 5 data sets with different control inputs are required to uniquely estimate the 36 parameters of this model.
Tables~\ref{tab:tsmn_num_of_data_sets} and~\ref{tab:tsmn_21_num_of_data_sets} in Appendix demonstrate the results of the numerical experiments when the number of data sets vary from 1 to 5,
and time measurements for each of the 8 components of the system is 240 and 20 respectively. Table~\ref{tab:tsmn_num_of_data_sets} demonstrates that when the number of data sets increases from 1 to 5 with accuracy $10^{-3}$, the number of identified parameters increases as 22, 27, 32, 34 and 36 accordingly, provided that 240 time measurements are given. Table~\ref{tab:tsmn_21_num_of_data_sets} demonstrates that with 20 time measurements the same number increases as 11, 24, 32, 33, 36.

\subsection{Range of convergence}\label{Range of convergence}
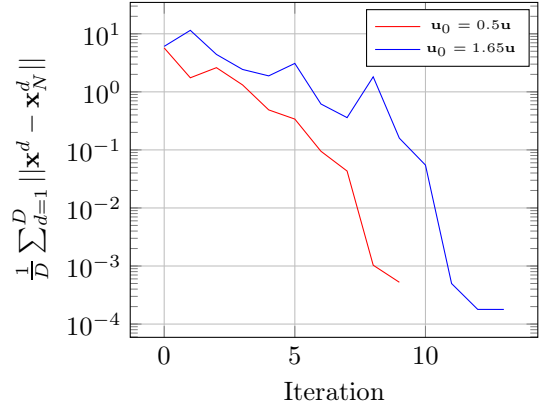
\begin{figure}
\begin{tikzpicture}
\begin{semilogyaxis}[
	xlabel=Iteration,
	ylabel=\(\frac{1}{D}\sum^D_{d=1}||\mathbf{x}^{d} - \mathbf{x}^{d}_{N}||\),
	grid=major,
	legend style={font=\tiny}
]
\addplot[color=red,grid=major] coordinates {
(0,5.70292407)
(1,1.74707669)
(2,2.60853879)
(3,1.31609525)
(4,0.488777551)
(5,0.339466404)
(6,0.0947009642)
(7,0.0432188813)
(8,0.00103234662)
(9,0.000524829663)
}; 
\addplot[color=blue,grid=major] coordinates {
(0,6.0742474)
(1,11.4680232)
(2,4.4109012)
(3,2.4300626)
(4,1.88452133)
(5,3.08908918)
(6,0.61754295)
(7,0.359846817)
(8,1.81497492)
(9,0.159231997)
(10,0.0547890382)
(11,0.000498489499)
(12,0.000178896808)
(13,0.000178695713)
}; 
\legend{\(\mathbf{u}_{0} = 0.5\mathbf{u}\),\(\mathbf{u}_{0} = 1.65\mathbf{u}\)}
\end{semilogyaxis}
\end{tikzpicture}
\caption{The evolution of the error, with 5 data sets, using Type I regularization, starting at varied \(\mathbf{u}_{0}\), \(t_0 = 0\), \(t_1 = 120\), \(\Delta t = 0.5\), i.e. 240 time measurements are given. Regularization parameter \(\alpha\) was chosen optimally.}
\label{fig:tsmn_no_noise_type1}
\end{figure}

We define the range of convergence as a neighborhood of the true solution $\mathbf{u}$ in $\mathbb{R}^{36}$ such that for any $\mathbf{u}_0$ chosen from it, the sequence $\mathbf{u}_N$ constructed according to our algorithm converges to $\mathbf{u}$. Consider the rectangular prism neighborhood of $\mathbf{u}$:
\[ \mathcal{P}_\tau^\omega=\{p\in \mathbb{R}^{36}: \tau u_i\leq p_i \leq \omega u_i, \ i=1,...,36\} \]
where $\tau$ and $\omega$ are two positive real numbers satisfying $\tau<1<\omega$. Numerical analysis demonstrates that for our model example, $\mathcal{P}_{0.5}^{1.65}$ is the largest rectangular prism contained in the convergence range according to the algorithm accompanied by Type I regularization. Figure~\ref{fig:tsmn_no_noise_type1} demonstrates the convergence with initial iteration $\mathbf{u}_0$ chosen at extremes of $\mathcal{P}_{0.5}^{1.65}$, namely $0.5 \mathbf{u}$ and $1.65 \mathbf{u}$, respectively.

\begin{figure}
\begin{tikzpicture}
\begin{semilogyaxis}[
	xlabel=Iteration,
	ylabel=\(\frac{1}{D}\sum^D_{d=1}||\mathbf{x}^{d} - \mathbf{x}^{d}_{N}||\),
	grid=major,
	legend style={font=\tiny}
]
\addplot[color=red,grid=major] coordinates {
(0,9.82297433)
(1,5.42329996)
(2,2.19255016)
(3,7.06950159)
(4,9.12882886)
(5,8.2315916)
(6,2.37389243)
(7,1.95000683)
(8,2.04035439)
(9,0.731771164)
(10,0.161493739)
(11,0.152875504)
(12,0.0105201554)
(13,0.000170046408)
(14,9.25315998e-05)
}; 
\addplot[color=blue,grid=major] coordinates {
(0,16.3919904)
(1,8.76320412)
(2,2.73484974)
(3,1.05615074)
(4,0.456010161)
(5,0.163442593)
(6,0.00269410596)
(7,0.000140516249)
}; 
\legend{\(\mathbf{u}_{0} = 0.03\mathbf{u}\),\(\mathbf{u}_{0} = 1001\mathbf{u}\)}
\end{semilogyaxis}
\end{tikzpicture}
\caption{The evolution of error with 5 data sets, using Type 2 regularization, starting at varied \(\mathbf{u}_{0}\), \(t_0 = 0\), \(t_1 = 120\), \(\Delta t = 0.5\), i.e. 240 time measurements are given. Regularization parameter \(\alpha\) was chosen optimally.}
\label{fig:tsmn_no_noise_type2}
\end{figure}
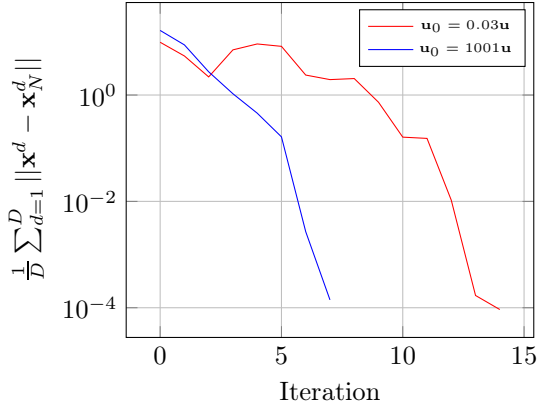

Careful implementation of Type II regularization allows significant expansion of the convergence range. In fact, by selecting $\mathbf{u}^*$ at the extremes of $\mathcal{P}_{0.5}^{1.65}$,
namely $\mathbf{u}^*=0.5\mathbf{u}$ and $\mathbf{u}^*=1.65\mathbf{u}$ we increased the convergence range to $\mathcal{P}_{0.03}^{\infty}$ according to the algorithm accompanied by Type II regularization. Figure~\ref{fig:tsmn_no_noise_type2} demonstrates the results of convergence of the method with Type II regularization when $\mathbf{u}^*=0.5\mathbf{u}$, and initial iteration is chosen as $0.03\mathbf{u}$. It also demonstrates the results when $\mathbf{u}^*=1.65\mathbf{u}$, and initial iteration is chosen as $1001\mathbf{u}$.

\subsection{Convergence with Noisy Measurements}\label{Convergence with Noisy Measurements}
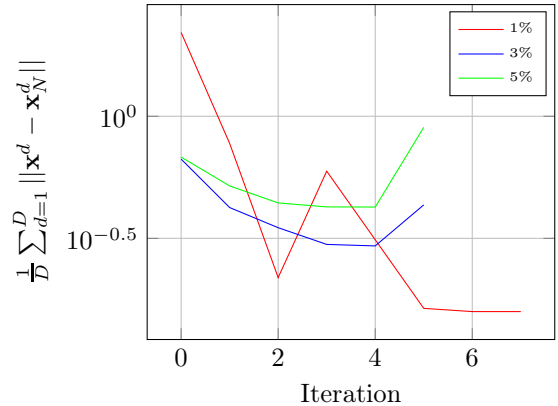
\begin{figure}
\begin{tikzpicture}
\begin{semilogyaxis}[
	xlabel=Iteration,
	ylabel=\(\frac{1}{D}\sum^D_{d=1}||\mathbf{x}^{d} - \mathbf{x}^{d}_{N}||\),
	grid=major,
	legend style={font=\tiny}
]
\addplot[color=red,grid=major] coordinates {
(0,2.20316863)
(1,0.773923373)
(2,0.218451316)
(3,0.59573087)
(4,0.312181724)
(5,0.163420005)
(6,0.158491009)
(7,0.158481913)
}; 
\addplot[color=blue,grid=major] coordinates {
(0,0.667629518)
(1,0.422749007)
(2,0.34962145)
(3,0.298647828)
(4,0.294433502)
(5,0.433455539)
}; 
\addplot[color=green,grid=major] coordinates {
(0,0.680640167)
(1,0.51841166)
(2,0.441604308)
(3,0.425369742)
(4,0.424751589)
(5,0.90079384)
}; 
\legend{\(1\%\), \(3\%\), \(5\%\)}
\end{semilogyaxis}
\end{tikzpicture}
\caption{The evolution of error for varied levels of noise using 5 data sets, \(t_0 = 0\), \(t_1 = 120\), \(\Delta t = 0.5\), i.e. 240 time measurements are given.}
\label{fig:tsmn_noise_combo}
\end{figure}

\begin{figure}
\begin{tikzpicture}
\begin{semilogyaxis}[
	xlabel=Iteration,
	ylabel=\(\frac{1}{D}\sum^D_{d=1}||\mathbf{x}^{d} - \mathbf{x}^{d}_{N}||\),
	grid=major,
	legend style={font=\tiny}
]
\addplot[color=red,grid=major] coordinates {
(0,2.21739559)
(1,0.944342685)
(2,0.406234316)
(3,0.335683577)
(4,0.314542193)
(5,0.314176087)
(6,0.314135215)
}; 
\addplot[color=blue,grid=major] coordinates {
(0,2.26458384)
(1,1.83549887)
(2,0.706905281)
(3,0.583524379)
(4,0.579134302)
(5,0.579018077)
}; 
\addplot[color=green,grid=major] coordinates {
(0,2.36050113)
(1,2.33604932)
(2,1.09003061)
(3,1.03150738)
(4,0.943826946)
(5,0.941525303)
(6,0.941484101)
(7,0.941444838)
(8,0.941407961)
(9,0.941377548)
}; 
\legend{\(1\%\), \(3\%\), \(5\%\)}
\end{semilogyaxis}
\end{tikzpicture}
\caption{The evolution of error for varied levels of noise using 5 data sets, \(t_0 = 0\), \(t_1 = 120\), \(\Delta t = 6.0\), i.e. 20 time measurements are given.}
\label{fig:tsmn_noise_21_combo}
\end{figure}
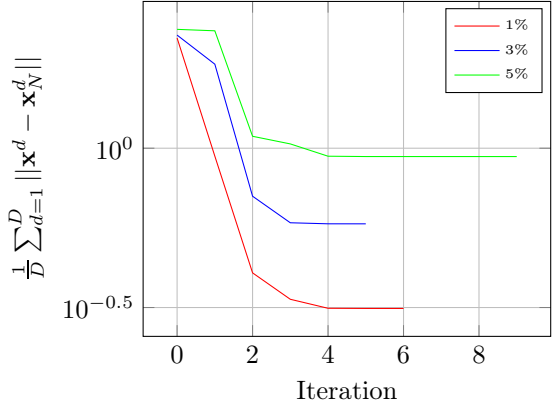

We pursued numerical experiments with simulated noisy data with Gaussian noise
\begin{equation}\label{noise}
y_i= x^d_i(t; \mathbf{u})+p x^d_i(t; \mathbf{u}) \nu_i, \ i=1,...,n
\end{equation}
where $p$ is a percentage and $\nu_i$ is a random variable with standard normal distribution:
\[ \nu_i \sim N(0,1). \]
Figure~\ref{fig:tsmn_noise_combo} demonstrates the convergence
in the experiment with 5 data sets and 240 noisy time measurements with $p=1, 3$ and \(5\).
In Figures~\ref{fig:tsmn_noise_boxplot_normdiff} and~\ref{fig:tsmn_noise_boxplot_paramdiff} we show the box plot based on 100 simulations for the residual and parameter vector error dependence on the noise percentage $p$.
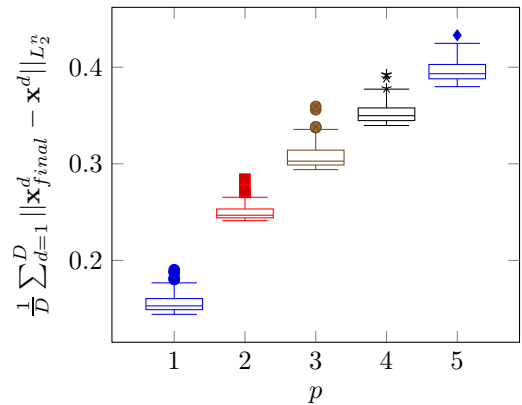
\begin{figure}
    \centering
    \begin{tikzpicture}
        \begin{axis}[
            boxplot/draw direction=y,
            xlabel=$p$,
            ylabel=$\frac{1}{D}\sum_{d=1}^D||\mathbf{x}^d_{final} - \mathbf{x}^d||_{L^n_2}$
        ]
        \addplot+[boxplot prepared={draw position=1,
                lower whisker=0.144173604,
                lower quartile=0.149015229,
                median=0.1529444955,
                upper quartile=0.1604848965,
                upper whisker=0.176758903
        }]
        table[row sep=\\,y index=0]{
                0.180304623\\ 0.180495223\\ 0.181853271\\ 0.18756463\\ 0.188218682\\ 0.190325565\\
        };
        coordinates {};
        \addplot+[boxplot prepared={draw position=2,
                lower whisker=0.241111149,
                lower quartile=0.244089382,
                median=0.246791328,
                upper quartile=0.253246228,
                upper whisker=0.265342318
        }]
        table[row sep=\\,y index=0]{
                0.270436234\\ 0.271912443\\ 0.284167737\\
        };
        coordinates {};
        \addplot+[boxplot prepared={draw position=3,
                lower whisker=0.294047719,
                lower quartile=0.29877835825,
                median=0.30280899250000004,
                upper quartile=0.3141686335,
                upper whisker=0.335616752
        }]
        table[row sep=\\,y index=0]{
                0.337387858\\ 0.338184393\\ 0.338285118\\ 0.355989581\\ 0.359399419\\
        };
        coordinates {};
        \addplot+[boxplot prepared={draw position=4,
                lower whisker=0.339647034,
                lower quartile=0.34475446425,
                median=0.34982134249999997,
                upper quartile=0.35785213275,
                upper whisker=0.377288515
        }]
        table[row sep=\\,y index=0]{
                0.378146243\\ 0.388711023\\ 0.392049693\\ 0.39266335\\
        };
        coordinates {};
        \addplot+[boxplot prepared={draw position=5,
                lower whisker=0.379901056,
                lower quartile=0.38801577100000006,
                median=0.393259143,
                upper quartile=0.4028554645,
                upper whisker=0.424751589
        }]
        table[row sep=\\,y index=0]{
                0.433107764\\
        };
        coordinates {};
        \end{axis}
    \end{tikzpicture}
    \caption{Distribution of the average residual error for several noise levels. Each run had 240 time measurements.}
    \label{fig:tsmn_noise_boxplot_normdiff}
\end{figure}
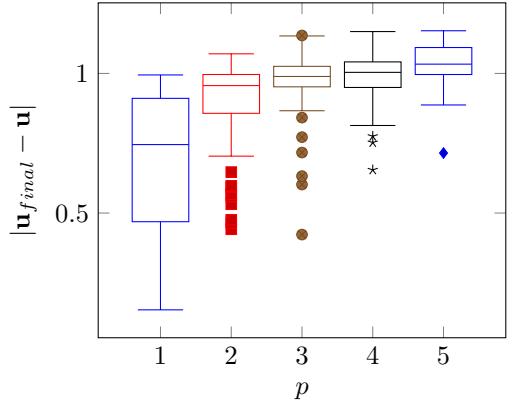
\begin{figure}
    \centering
    \begin{tikzpicture}
        \begin{axis}[
            boxplot/draw direction=y,
            xlabel=$p$,
            ylabel=$|\mathbf{u}_{final} - \mathbf{u}|$
        ]
        \addplot+[boxplot prepared={draw position=1,
                lower whisker=0.153271681,
                lower quartile=0.46888671425,
                median=0.7451239510000001,
                upper quartile=0.91076298125,
                upper whisker=0.994563629
        }]
        table[row sep=\\,y index=0]{
                \\
        };
        coordinates {};
        \addplot+[boxplot prepared={draw position=2,
                lower whisker=0.703705879,
                lower quartile=0.8570780785,
                median=0.956436953,
                upper quartile=0.9961812620000001,
                upper whisker=1.07020866
        }]
        table[row sep=\\,y index=0]{
                0.441910485\\ 0.464372297\\ 0.475772237\\ 0.531433012\\ 0.5594125\\ 0.567340025\\ 0.593846342\\ 0.597115432\\ 0.646811644\\
        };
        coordinates {};
        \addplot+[boxplot prepared={draw position=3,
                lower whisker=0.866336533,
                lower quartile=0.95199610675,
                median=0.989056028,
                upper quartile=1.02532268,
                upper whisker=1.13425145
        }]
        table[row sep=\\,y index=0]{
                1.13589268\\ 0.422818949\\ 0.602003712\\ 0.632195054\\ 0.716642222\\ 0.771830049\\ 0.841749787\\
        };
        coordinates {};
        \addplot+[boxplot prepared={draw position=4,
                lower whisker=0.813534696,
                lower quartile=0.94974315425,
                median=1.003843055,
                upper quartile=1.0412687025,
                upper whisker=1.14990558
        }]
        table[row sep=\\,y index=0]{
                0.654534059\\ 0.75243191\\ 0.77501399\\
        };
        coordinates {};
        \addplot+[boxplot prepared={draw position=5,
                lower whisker=0.887036155,
                lower quartile=0.99610049225,
                median=1.03322002,
                upper quartile=1.0927107075,
                upper whisker=1.15279507
        }]
        table[row sep=\\,y index=0]{
                0.714894508\\
        };
        coordinates {};
        \end{axis}
    \end{tikzpicture}
    \caption{Distribution of the parameter error at several noise levels. Each run had 240 time measurements.}
    \label{fig:tsmn_noise_boxplot_paramdiff}
\end{figure}
Similar results
with 20 noisy time measurements are given in Figure~\ref{fig:tsmn_noise_21_combo}.

\subsection{Rate of convergence}\label{cr}
To estimate the convergence rate $\gamma$ from the relation
\[
|\mathbf{u}_{k+1} - \mathbf{u}_{k}| \sim C|\mathbf{u}_k - \mathbf{u}_{k-1}|^\gamma
\]
we plot $\log|\mathbf{u}_{k+1} - \mathbf{u}_{k}|$ vs. $\log|\mathbf{u}_{k} - \mathbf{u}_{k-1}|$ and find a line of best fit to identify $\gamma$ and $C$. Figures~\ref{fig:conv_rate_16} and~\ref{fig:conv_rate_05} demonstrate the outcome. For the numerical experiment for the green line in  Figure~\ref{fig:tsmn_no_noise_combo} we have  $\gamma=1.6104, C = 3.1622E-3$, and for the black line we have $\gamma=1.1674, C = 6.3271E-1$. The difference in convergence rate of two examples is in particular due to choice of the regularization parameter
$\alpha$. Choosing the optimal choice for $\alpha$, as it is demonstrated in  Figure~\ref{fig:tmsn01_alpha_all} vs. Figure~\ref{fig:tmsn02_alpha_all}, causes higher convergence rate for the numerical experiment expressed in  Figure~\ref{fig:conv_rate_16} vs. Figure~\ref{fig:conv_rate_05}. We expect theoretical convergence rate of the method is quadratic \cite{abdulla1}.
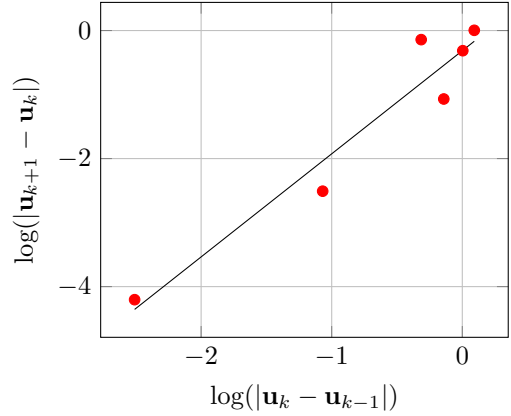
\begin{figure}
\begin{tikzpicture}
\begin{axis}[
	xlabel=\(\log(|\mathbf{u}_{k} - \mathbf{u}_{k-1}|)\),
	ylabel=\(\log(|\mathbf{u}_{k+1} - \mathbf{u}_{k}|)\),
	grid=major,
	legend style={font=\tiny}
]
\addplot[color=red,
	grid=major,
	mark = *,
	only marks
	] coordinates {
	(0.0917, 0.0038)
	(0.0038, -0.3147)
	(-0.3147,-0.1418)
	(-0.1418,-1.0691)
	(-1.0691,-2.5092)
	(-2.5092,-4.2040)
};
\addplot[color = black,
	mark  = none
	] coordinates {
	(9.174460e-02,-1.674937e-01)
	(-2.509182e+00,-4.355866e+00)
};
\end{axis}
\end{tikzpicture}
\caption{The convergence rate graph corresponding to green curve in Figure~\ref{fig:tsmn_no_noise_combo}, where \(r = 1.6104\) and \(C = 3.1622E-3\).}
\label{fig:conv_rate_16}
\end{figure}
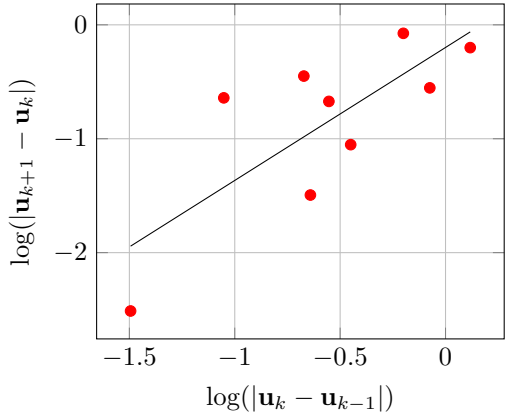
\begin{figure}
\begin{tikzpicture}
\begin{axis}[
	xlabel=\(\log(|\mathbf{u}_{k} - \mathbf{u}_{k-1}|)\),
	ylabel=\(\log(|\mathbf{u}_{k+1} - \mathbf{u}_{k}|)\),
	grid=major,
	legend style={font=\tiny}
]
\addplot[color=red,
    grid=major,
    mark  = *,
    only marks
    ] coordinates {
    (0.1172, -0.2002)
(-0.2002,-0.0746)
(-0.0746,-0.5534)
(-0.5534,-0.6722)
(-0.6722,-0.4499)
(-0.4499,-1.0520)
(-1.0520,-0.6411)
(-0.6411,-1.4942)
(-1.4942,-2.5123)
};
\addplot[color = black,mark  = none] coordinates {
(1.171605e-01,-6.201818e-02)
(-1.494164e+00,-1.943071e+00)
};
\end{axis}
\end{tikzpicture}
\caption{The convergence rate graph corresponding to black curve in Figure~\ref{fig:tsmn_no_noise_combo}, where \(r = 1.1674\) and \(C = 6.3271E-1\).}
\label{fig:conv_rate_05}
\end{figure}

\subsection{Convergence with Partial Measurements}\label{Convergence with Partial Measurements}
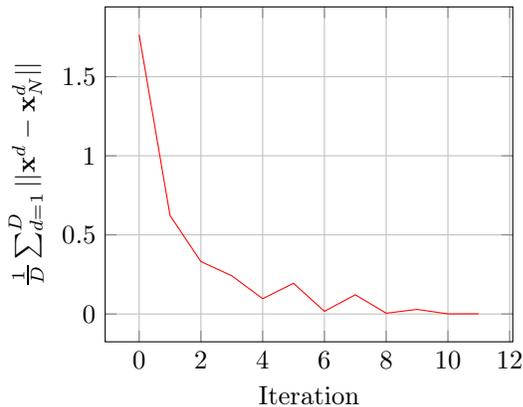
\begin{figure}
\begin{tikzpicture}
\begin{axis}[
	xlabel=Iteration,
	ylabel=\(\frac{1}{D}\sum^D_{d=1}||\mathbf{x}^{d} - \mathbf{x}^{d}_{N}||\),
	grid=major,
	legend style={font=\tiny}
]
\addplot[color=red,grid=major] coordinates {
(0,1.76497443)
(1,0.622306232)
(2,0.331343075)
(3,0.241340147)
(4,0.09663908)
(5,0.19375643)
(6,0.0167733352)
(7,0.121303422)
(8,0.00472259583)
(9,0.0289878331)
(10,0.000499476745)
(11,0.000880992746)
};
\end{axis}
\end{tikzpicture}
\caption{The average error at each iteration for 5 data sets with partial measurements given for 4 out of 8 states, \(t_0 = 0\), \(t_1 = 120\), \(\Delta t = 6.0\), i.e. 20 time measurements for 4 components are given. Regularization parameter \(\alpha\) was determined using \eqref{eq:alphaformula} where \(C = 0.5\) and \(\gamma = 2\).}
\label{fig:partial_tsmn02}
\end{figure}

We tested the convergence of the method when only some of the components of the system have available measurements or partial measurements. In this case the inverse problem must be solved with partial observations. A typical result is demonstrated in Figure~\ref{fig:partial_tsmn02}. We considered our numerical experiment with 5 data sets, and with 20 time measurements of only components 3, 4, 5, and 7. Figure~\ref{fig:partial_tsmn02} demonstrates the convergence, although convergence rate slowed down in comparison with the experiment when full set of measurements are given.

\subsection{Comparison with {\it lsqnonlin}, {\it nl2sol} and {\it fmincon}}\label{comparison}
As in our previous paper \cite{previouspaper} we are comparing our method {\it qlopt} with the most popular methods available as open software \cite{amigo2,data2dynamics,pesto}
such as
\begin{itemize}
\item Levenberg-Marquardt algorithm and trust-region-reflective method (function {\it lsqnonlin} in MatLab) \cite{hanke1997regularizing}.
\item An adaptive non-linear least-squares algorithm (function {\it nl2sol} in MatLab) \cite{nl2sol}.
\item Sequential Quadratic Programming (function {\it fmincon} in MatLab Optimization toolbox)
\end{itemize}
We used model example provided by AMIGO2, which had 16 data sets with each component evaluated at 21 time points giving a total of 2688 data points.
We ran each algorithm 20 times and recorded the {\it average relative error of the parameter values (r.e.), the median number of objective function evaluations (f.e.),
the average computational time (c.t.), and the median number of iterations (n.i.)}.  The results are demonstrated in Table \ref{tab:tsmn_comparison}. Algorithm performed by {\it fmincon} did not converge. All three other methods have a comparable relative error. In terms of required number of iterations, our method is comparable to {\it nl2sol}, and both have a clear advantage over {\it lsqnonlin}. In terms of {\it computational time} and {\it function evaluations} our method has an enormous advantage over both methods. It should be noted that our software package {\ qlopt} is using C++ and Eigen, which gives an advantage over MatLab-based methods with respect to computational time.
\begin{table}
	\sisetup{
	    round-mode = places,
	    round-precision = 3,
		tight-spacing = true,
	    table-number-alignment = center
	}
	\begin{tabularx}{\linewidth}{lSSSS}
		\toprule
			metric		& {qlopt}  	 			& {lsqnonlin}       & {nl2sol}	 \\ \midrule
			r.e.		& \num{4.27116236e-4} 	& \num{3.2356e-5}   & \num{4.3538e-4}		\\ \midrule
			n.i.		& 8		 				& 16	            & 7 	         \\ \midrule
			c.t. (s) 	& 1.3887635       		& 7.6563		    & 5.0426	     \\ \midrule
			f.e.		& 8        				& 593			    & 299	     \\ \midrule

		\bottomrule
	\end{tabularx}
\caption{Comparison of several local optimization methods with the
presented method for the three step metabolic network.
The relative errors (r.e.),
the number of iterations (n.i.),
mean of the computational time (c.t.) of 20 runs,
and the number of function evaluations (f.e.) are compared in numerical experiment designed by AMIGO2 which contained 16 datasets with 21 time measurements per component and per data set.}
\label{tab:tsmn_comparison}
\end{table}


\section{Benchmark Model 2: Central Carbon Metabolism of Escherichia coli}\label{B2}
The method is tested in a benchmark model of central carbon metabolism of {\it Escherichia coli} introduced in~\cite{b2model} and labeled as model B2 in~\cite{benchmarkpaper}. The mathematical model reproduces the response to a pulse in extracellular glucose concentration. It consists of a system of 18 differential equations for the concentrations of metabolites, which include 17 intracellular metabolites and extracellular glycose. As in ~\cite{benchmarkpaper} we consider inverse problem on the identification of 116 parameters which express kinetic properties and maximum reaction rates.

\subsection{Numerical Results with Noise-free Data}\label{b2nonoise}
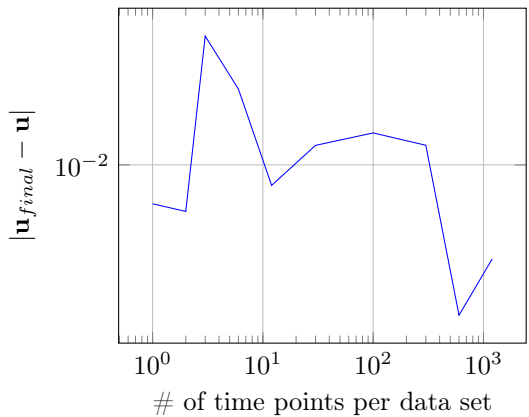
\begin{figure}
\begin{tikzpicture}
\begin{loglogaxis}[
xlabel=\# of time points per data set,
ylabel=\(|\mathbf{u}_{final} - \mathbf{u}|\),
grid=major,
legend style={font=\tiny}
]
\addplot[color=blue,grid=major] coordinates {
(1,0.0066435138)
(2,0.00613074064)
(3,0.038742303)
(6,0.022208027)
(12,0.00805980849)
(30,0.0122784587)
(100,0.0139972899)
(300,0.0122925017)
(600,0.00205832391)
(1200,0.00372218909)
};
\end{loglogaxis}
\end{tikzpicture}
\caption{Graph of the number time points vs.
\(|\mathbf{u}_{final} - \mathbf{u}|\) for benchmark model 2;  
$\mathbf{u}_0 = 1.75\mathbf{u}$, regularization parameter \(\alpha\) is chosen optimally.}
\label{fig:b2_data_pnts}
\end{figure}

\begin{figure}
\begin{tikzpicture}
\begin{loglogaxis}[
xlabel=\# of time points per data set,
ylabel=\(|\log(\mathbf{u}_{final}) - \log(\mathbf{u})|\),
grid=major,
legend style={font=\tiny}
]
\addplot[color=blue,grid=major] coordinates {
(1,3.63692859e-08)
(2,3.11279153e-08)
(3,1.43169261e-07)
(6,5.99067881e-08)
(12,2.70939945e-08)
(30,8.93898746e-08)
(100,6.00823614e-08)
(300,5.93241962e-08)
(600,2.2188427e-08)
(1200,4.70413021e-08)
};
\end{loglogaxis}
\end{tikzpicture}
\caption{Graph of the number time points vs.
\(|\log(\mathbf{u}_{final}) - \log(\mathbf{u})|\) for benchmark model 2; 
$\mathbf{u}_0 = 1.75\mathbf{u}$, regularization parameter \(\alpha\) is chosen optimally.}
\label{fig:b2_data_pnts_log}
\end{figure}
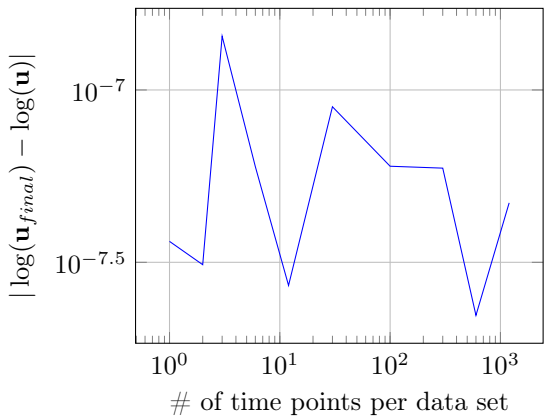

We applied the numerical method to identify the 116 parameters with one data set ($d=1$). We generated simulated measurements by solving the system of 18 linear ODEs with true values of 116 parameters. We choose the number of time data points for each of the 18 components of the system at 1200 uniformly distributed time grid points in the segment $[0, 300]$. Note that since the system is linear, there is no need on quasilinearization step in our algorithm. Computational cost of each iteration consists of solution of the original system of 18 linear ODEs; solving a system of 2088 linear ODEs to identify sensitivity matrix-function; calculation of 13572 integrals for entries of the matrix $A_N$ and vector $P_N$; and finally solving a  system of 116 linear algebraic equations to find the increment of the parameters. 
We applied our algorithm by selecting number of uniformly distributed in $[0, 300]$ data points varying between 1 and 1200.
The method is extremely robust and convergence is still the case if the number of data points is reduced to a single time measurement at the end of the time interval for each of the
18 components of the system. Figure~\ref{fig:b2_data_pnts} demonstrates the dependence of the number of time measurements for each component on the Euclidean norm difference of the true parameter vector from the parameter vector at the final iteration. 
Figure~\ref{fig:b2_data_pnts_log} demonstrate the results of the same experiment pursued in logarithmic scale. In general, solving problems in the logarithmic scale improves convergence, especially when the parameters vary greatly in magnitude~\cite{survey2, raue2013lessons}.

Next, we pursued an experiment with partial measurements:  100 data points taken only for the 9 components of the system. This is similar to experiments analyzed in~\cite{b2model, benchmarkpaper}. Figure~\ref{fig:b2_no_noise} demonstrates rapid numerical convergence. Error $||\mathbf{x} - \mathbf{x}_{N}||_{L_2^n}$ reaches desired accuracy $10^{-5}$ in 8 steps, if initial iteration is chosen as $\bf{u}_0=1.75  \bf{u}$, where $\bf{u}$ is a true parameter vector. Figure~\ref{fig:b2_conv_rate_no_noise} demonstrates that the algorithm maintains superlinear convergence rate.
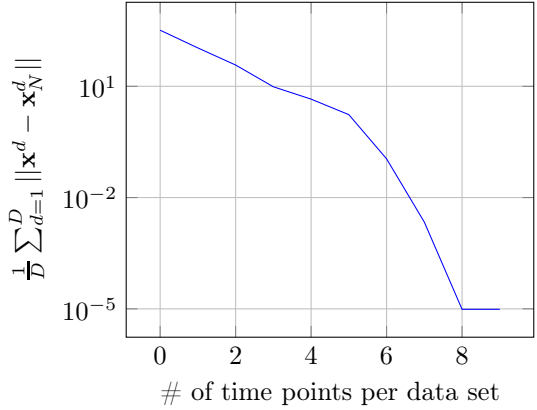
\begin{figure}
\begin{tikzpicture}
\begin{semilogyaxis}[
xlabel=\# of time points per data set,
ylabel=\(\frac{1}{D}\sum^D_{d=1}||\mathbf{x}^{d} - \mathbf{x}^{d}_{N}||\),
grid=major,
legend style={font=\tiny}
]
\addplot[color=blue,grid=major] coordinates {
(0,325.200209)
(1,108.104435)
(2,37.448763)
(3,9.69700375)
(4,4.50847952)
(5,1.71975017)
(6,0.110752297)
(7,0.00218178989)
(8,9.73389964e-06)
(9,9.73389964e-06)
};
\end{semilogyaxis}
\end{tikzpicture}
\caption{Graph of
\(\frac{1}{D}\sum^D_{d=1}||\mathbf{x}^{d} - \mathbf{x}^{d}_{N}||\) for benchmark model 2 with partial measurement without noise;
$\mathbf{u}_0 = 1.75\mathbf{u}$.}
\label{fig:b2_no_noise}
\end{figure}
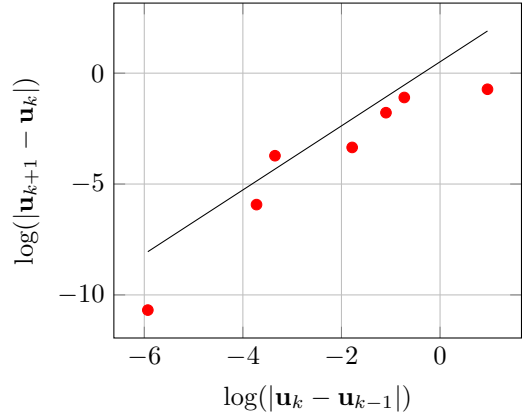
\begin{figure}
\begin{tikzpicture}
\begin{axis}[
	xlabel=\(\log(|\mathbf{u}_{k} - \mathbf{u}_{k-1}|)\),
	ylabel=\(\log(|\mathbf{u}_{k+1} - \mathbf{u}_{k}|)\),
	grid=major,
	legend style={font=\tiny}
]
\addplot[color=red,
    grid=major,
    mark  = *,
    only marks
    ] coordinates {
		(0.962403979154,-0.725604597569)
		(-0.725604597569,-1.09702752707)
		(-1.09702752707,-1.7829633686)
		(-1.7829633686,-3.34905956713)
		(-3.34905956713,-3.7231312754)
		(-3.7231312754,-5.92789194732)
		(-5.92789194732,-10.6841760949)
	};
\addplot[color = black,mark  = none] coordinates {
	(0.962403979154,1.90148989054)
	(-5.92789194732,-8.05094885934)
};
\end{axis}
\end{tikzpicture}
\caption{The convergence rate graph corresponding to Figure~\ref{fig:b2_no_noise}, with \(r = 1.4444\) and \(C = 0.5114\).}
\label{fig:b2_conv_rate_no_noise}
\end{figure}

\subsection{Convergence with Noisy Measurements}\label{b2noise}
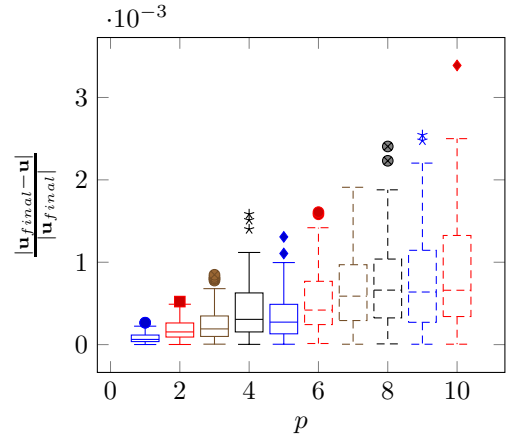
\begin{figure}
    \centering
    \begin{tikzpicture}
        \begin{axis}[
            boxplot/draw direction=y,
            xlabel=$p$,
            ylabel=$\frac{|\mathbf{u}_{final} - \mathbf{u}|}{|\mathbf{u}_{final}|}$
        ]
        \addplot+[boxplot prepared={draw position=1,
                lower whisker=1.10895549577e-06,
                lower quartile=3.67781402042e-05,
                median=6.32160507651e-05,
                upper quartile=0.000116203312072,
                upper whisker=0.000223826783477
        }]
        table[row sep=\\,y index=0]{
                0.000264991957741\\ 0.000266749305901\\
        };
        coordinates {};
        \addplot+[boxplot prepared={draw position=2,
                lower whisker=1.50730321615e-06,
                lower quartile=9.18360549267e-05,
                median=0.000153986350977,
                upper quartile=0.000262832232413,
                upper whisker=0.000490216824509
        }]
        table[row sep=\\,y index=0]{
                0.000522309140074\\
        };
        coordinates {};
        \addplot+[boxplot prepared={draw position=3,
                lower whisker=5.555134157e-06,
                lower quartile=9.9261474094e-05,
                median=0.000190742039858,
                upper quartile=0.000347899190699,
                upper whisker=0.00067938884262
        }]
        table[row sep=\\,y index=0]{
                0.000777258897436\\ 0.000796753743394\\ 0.000803730500415\\ 0.000821656270606\\ 0.000848121790883\\
        };
        coordinates {};
        \addplot+[boxplot prepared={draw position=4,
                lower whisker=3.01668588827e-06,
                lower quartile=0.000153999112348,
                median=0.00030654663362,
                upper quartile=0.000627533876142,
                upper whisker=0.00111859575999
        }]
        table[row sep=\\,y index=0]{
                0.001400510917\\ 0.0015115284036\\ 0.00158394643269\\
        };
        coordinates {};
        \addplot+[boxplot prepared={draw position=5,
                lower whisker=4.6110208356e-06,
                lower quartile=0.000131058046604,
                median=0.000272987369136,
                upper quartile=0.00048911980606,
                upper whisker=0.00099603125111
        }]
        table[row sep=\\,y index=0]{
                0.00110610577335\\ 0.00130765424441\\
        };
        coordinates {};
        \addplot+[boxplot prepared={draw position=6,
                lower whisker=1.34172921324e-05,
                lower quartile=0.000243247790967,
                median=0.000419271304743,
                upper quartile=0.000768502058959,
                upper whisker=0.00141911227739
        }]
        table[row sep=\\,y index=0]{
                0.0015818368368\\ 0.00160838171998\\
        };
        coordinates {};
        \addplot+[boxplot prepared={draw position=7,
                lower whisker=4.76041673817e-06,
                lower quartile=0.000292232946921,
                median=0.000588294204572,
                upper quartile=0.000970025716902,
                upper whisker=0.00190993451365
        }]
        table[row sep=\\,y index=0]{
                \\
        };
        coordinates {};
        \addplot+[boxplot prepared={draw position=8,
                lower whisker=8.90110179879e-06,
                lower quartile=0.000324273300603,
                median=0.000661530019449,
                upper quartile=0.00103855037133,
                upper whisker=0.00187909880705
        }]
        table[row sep=\\,y index=0]{
                0.00223129738607\\ 0.00240550047697\\
        };
        coordinates {};
        \addplot+[boxplot prepared={draw position=9,
                lower whisker=4.61712965121e-06,
                lower quartile=0.000270722121199,
                median=0.000638476151086,
                upper quartile=0.00114478838656,
                upper whisker=0.00220331593016
        }]
        table[row sep=\\,y index=0]{
                0.00247719324698\\ 0.00254451219243\\
        };
        coordinates {};
        \addplot+[boxplot prepared={draw position=10,
                lower whisker=5.13021899911e-06,
                lower quartile=0.000340943593004,
                median=0.000659943323567,
                upper quartile=0.00132490579629,
                upper whisker=0.00250030998088
        }]
        table[row sep=\\,y index=0]{
                0.0033890805986\\
        };
        coordinates {};
        \end{axis}
    \end{tikzpicture}
    \caption{Distribution of the relative parameter error for the benchmark model 2 at several noise levels. Each run had 100 data points.}
    \label{fig:rel_noise_boxplot_relparamdiff}
\end{figure}

We pursued numerical experiments with simulated noisy data with Gaussian noise as in \eqref{noise}.
Figure~\ref{fig:rel_noise_boxplot_relparamdiff} demonstrates the box plot based on 100 simulations for the parameter vector error dependence on the noise percentage $p$ changing between 1\% to 10\%. In every experiment the iterative value of the regularization parameter $\alpha$ was chosen optimally.

\subsection{Range of convergence}\label{b2rc}
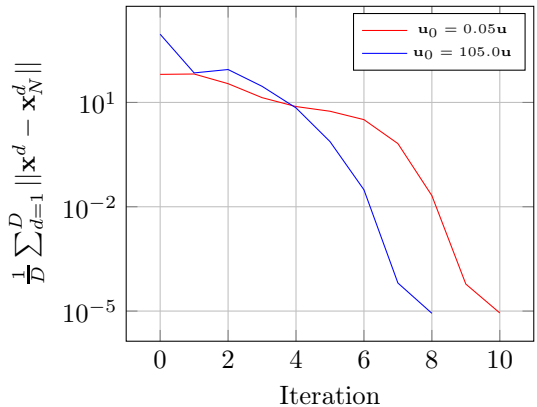
\begin{figure}
\begin{tikzpicture}
\begin{semilogyaxis}[
	xlabel=Iteration,
	ylabel=\(\frac{1}{D}\sum^D_{d=1}||\mathbf{x}^{d} - \mathbf{x}^{d}_{N}||\),
	grid=major,
	legend style={font=\tiny}
]
\addplot[color=red,grid=major] coordinates {
(0,64.1424358)
(1,65.7267728)
(2,34.639746)
(3,13.7359933)
(4,7.61321625)
(5,5.582004)
(6,3.20155291)
(7,0.654969669)
(8,0.0208897296)
(9,5.97158547e-05)
(10,8.82316192e-06)
}; 
\addplot[color=blue,grid=major] coordinates {
(0,916.062072)
(1,70.6427061)
(2,88.3715137)
(3,29.1079594)
(4,6.96472497)
(5,0.749800119)
(6,0.0307682399)
(7,6.41185459e-05)
(8,8.61927893e-06)
}; 
\legend{\(\mathbf{u}_{0} = 0.05\mathbf{u}\),\(\mathbf{u}_{0} = 105.0\mathbf{u}\)}
\end{semilogyaxis}
\end{tikzpicture}
\caption{The evolution of error using type 1 regularization, starting at varied \(\mathbf{u}_{0}\), \(t_0 = 0\), \(t_1 = 300\), \(\Delta t = 3.0\), i.e. 100 time points. Regularization parameter \(\alpha\) was chosen optimally.}
\label{fig:b2_no_noise_type1}
\end{figure}

Numerical analysis demonstrates that for our model example B2 using the logarithmic scale for the parameters, with the help of Type II regularization, the convergence range can be extended to $\mathcal{P}_{0.05}^{105}$ (before log transformation).   Figure~\ref{fig:b2_no_noise_type1} demonstrates the convergence with initial iteration $\mathbf{u}_0$ chosen at extremes of $\mathcal{P}_{0.05}^{105}$, namely $0.05 \mathbf{u}$ and $105 \mathbf{u}$, respectively.

\section{Conclusions}\label{conclusions}
This paper develops the numerical method for solving  inverse problem on the identification of parameters for large scale kinetic models in systems biology. The iterative method combines ideas of Pontryagin optimization or Bellman quasilinearization with sensitivity analysis and Tikhonov regularization. Embedding a method of staggered corrector for sensitivity analysis and by enhancing multi-objective optimization enables application of the method to large-scale models with practically non-identifiable
parameters based on multiple data sets, possibly with partial and noisy measurements. The method is tested in two benchmark kinetic models, such as three-step pathway modeled by 8 nonlinear ODEs with 36 unknown and two control input parameters, and 
a model of central carbon metabolism of {\it Escherichia coli} described by a system of 18 linear ODEs with 116 unknown parameters. Extensive analysis demonstrates that the modified method is extremely well adapted to large scale problems. The main conclusions of the paper can be summarized as follows:
\begin{itemize}
\item There is a minimum number of data sets with different control parameter inputs required to achieve superlinear/quadratic convergence and unique identifiability of parameters for large-scale problems.
\item Increase of data sets beyond the minimum doesn't significantly affect convergence rate and accuracy, but possibly affects the computational cost.
\item The method is extremely robust in terms of required number of time measurements for components of the system for every data set. For the benchmark models, high accuracy is achieved even with the single measurement for each component at the final time instance. \item Optimal choice of the Tikhonov regularization parameter significantly increases the convergence rate and precision.
\item The method is robust with respect to noisy measurements. Simulating up to 5\% Gaussian noise in benchmark models does not affect the convergence rate, but only adds some additional error to final output in accordance with the noise level.
\item Implementation of the Type II Tikhonov regularization significantly increases the convergence range of the algorithm.
\item Method is robust with respect to partial measurements. Application to the benchmark model with measurements of only half of the components demonstrates convergence with slightly reduced but still quite high accuracy.
\item The method is highly competitive and has an advantage over popular methods such as {\it lsqnonlin, fmincon, nl2sol} in terms of computational time, number of iterations and function evaluations.
\item Combination of the method with "multi-start" strategy and latin hypercube sampling will open a way to develop a powerful global optimization method with least computational cost, which is robust with respect to nonlinearities and scales well with problem size 
\end{itemize}

\textbf{References}
\bibliographystyle{elsarticle-num}
\bibliography{references}
\section{Appendix}\label{appendix}
Tables~\ref{tab:tsmn_num_of_data_sets} and~\ref{tab:tsmn_21_num_of_data_sets} demonstrate the dependence of the parameter identification on the number of data sets 1-5 for the Benchmark Model 1 (see Section~\ref{Convergence vs. Number of Data Sets}).
\begin{table}
\centering
\begin{adjustbox}{clip=false}
	\begin{tabular}{lSSS SSc}
		\toprule
		$\vecu$ & 1 &  2 &  3 &  4 & 5 & $\vecu$  \\ \midrule
		\(p_{1}\)  		& 0.993586455 		& 	0.99453233 			& 	1.00447944 			& 	0.999085904 			& 	1.00020433 				& 1 \\
		\(p_{2}\)  		& 1.24979552 		& 	1.23814882 			& 	1.19979403 			& 	0.873692785 			& 	1.00022216 				& 1 \\
		\(p_{3}\)  		& 2.4996709 			& 	2.49266164 			& 	2.46999011 			& 	2.34285705 				& 	1.99951318 				& 2 \\
		\(p_{4}\)  		& 1.25353032 		& 	1.31690074 			& 	1.14606682 			& 	1.00296597 				& 	0.999986751 			& 1 \\
		\(p_{5}\)  		& 2.4963291 			& 	2.48044375 			& 	1.89146594 			& 	1.99534059 				& 	1.99999288 				& 2 \\
		\(p_{6}\)  		&  1.00008034 		& 	1.00008094 			& 	1.0000297 			& 	1.00017986 				& 	1.0002031 				& 1 \\
		\(p_{7}\)  		& 0.999651067 		& 	0.997762842 		& 	0.999232944			& 	1.0000865 				& 	1.0004875 				& 1 \\
		\(p_{8}\)  		& 1.24971111 		& 	0.877819402 		& 	0.999838041			& 	0.999928742 			& 	0.999968907 			& 1 \\
		\(p_{9}\)  		& 2.49953505 		& 	2.3097861 			& 	1.99949878 			& 	2.00018473 				& 	2.00009361 				& 2 \\
		\(p_{10}\)		& 0.999512223 		& 	0.999134556 		& 	0.999864398			& 	0.999998913 			& 	0.999999413 			& 1 \\
		\(p_{11}\)		& 2.00149681 		& 	1.99936747 			& 	2.00013584 			& 	1.99982176 				& 	1.9998918 				& 2 \\
		\(p_{12}\)		& 1.00187965 		& 	0.998905034 		& 	0.999258494			& 	1.00007148 				& 	1.00047922 				& 1 \\
		\(p_{13}\)		& 0.998556816 		& 	1.00409558 			& 	0.997965171			& 	1.00149337 				& 	1.00171945 				& 1 \\
		\(p_{14}\)		& 1.24973112 		& 	0.880848321 		& 	0.999886368			& 	1.00018617 				& 	1.00005545 				& 1 \\
		\(p_{15}\)		& 2.49956726 		& 	2.30092875 			& 	1.99960801 			& 	1.99943668 				& 	1.99988298 				& 2 \\
		\(p_{16}\)		& 0.999137801 		& 	1.00139005 			& 	1.00012745 			& 	0.999519767 			& 	0.999872548 			& 1 \\
		\(p_{17}\)		& 1.99958513 		& 	1.99593174 			& 	1.99958681 			& 	2.00079195 				& 	2.00019573 				& 2 \\
		\(p_{18}\)	    & 1.00052325 		& 	1.00337571 			& 	0.997824575			& 	1.00194374 				& 	1.00184079 				& 1 \\
		\(p_{19}\)		& 0.0999752469 	& 	0.100010274 		& 	0.100005577			& 	0.100006903 			& 	0.0999975188				& 0.1 \\
		\(p_{20}\)		& 0.999669749 		& 	1.00004006 			& 	1.00008123 			& 	1.00010508 				& 	1.00006192 				& 1 \\
		\(p_{21}\)		& 0.0999917364 	& 	0.100008235 		& 	0.10000121 			& 	0.100001058 			& 	0.0999940189				& 0.1 \\
		\(p_{22}\)		& 0.100035979 		& 	0.100012274 		& 	0.100010336			& 	0.100006781 			& 	0.100001411 			& 0.1 \\
		\(p_{23}\)		& 1.00042909 		& 	1.00001885 			& 	0.999978182			& 	0.999996904 			& 	1.0000288 				& 1 \\
		\(p_{24}\)		& 0.100013807 		& 	0.10001135 			& 	0.100011834			& 	0.10000702 				& 	0.0999997043			& 0.1 \\
		\(p_{25}\)		& 0.100005293 		& 	0.100021289 		& 	0.10002808 			& 	0.099982281 			& 	0.0999975974			& 0.1 \\
		\(p_{26}\)		& 0.999911817 		& 	0.999930186 		& 	0.999950739			& 	0.99996916 				& 	0.999960958 			& 1 \\
		\(p_{27}\)		& 0.100011018 		& 	0.100026183 		& 	0.100031463			& 	0.0999846931			& 	0.100000705 			& 0.1 \\
		\(p_{28}\)		& 1.02525258 		& 	1.0000645 			& 	0.999922142			& 	1.00002979 				& 	0.999993611 			& 1 \\
		\(p_{29}\)		& 1.27901919 		& 	1.00028082 			& 	0.999627162			& 	0.999745336 			& 	1.00010507 				& 1 \\
		\(p_{30}\)		& 1.24744166 		& 	0.999797398 		& 	0.999437581			& 	0.999402741 			& 	1.00025053 				& 1 \\
		\(p_{31}\)		& 1.00035658 		& 	1.00022626 			& 	0.999358374			& 	0.999973262 			& 	1.00003599 				& 1 \\
		\(p_{32}\)		& 1.00123519 		& 	0.999794087 		& 	1.00099456 			& 	0.999792997 			& 	1.00000591 				& 1 \\
		\(p_{33}\)		& 1.00000069 		& 	0.998142174 		& 	1.00436027 			& 	0.999562622 			& 	0.999941151 			& 1 \\
		\(p_{34}\)		& 1.00035765 		& 	0.999804847 		& 	1.0001006 			& 	0.999900398 			& 	1.00006984 				& 1 \\
		\(p_{35}\)		& 1.01093863 		& 	0.99973188 			& 	1.00052432 			& 	0.999862607 			& 	1.00010126 				& 1 \\
		\(p_{36}\)		& 1.2590738 			& 	0.999286301 		& 	1.00097748 			& 	0.999973157 			& 	1.00008351 				& 1 \\
		\bottomrule
	\end{tabular}
\end{adjustbox}
\caption{The evolution of the parameters vs. the number of data sets changing from 1 to 5, with 
\(t_0 = 0\), \(t_1 = 120\), \(\Delta t = 0.5\), 240 time measurements for each of the 8 components are given, \(\mathbf{u}_0 = 1.25\mathbf{u} \), and \(\alpha\) is determined using \eqref{eq:alphaformula}.}
\label{tab:tsmn_num_of_data_sets}
\end{table}

\begin{table}
\centering
\begin{adjustbox}{clip=false}
	\begin{tabular}{lSSS SSc}
		\toprule
		$\vecu$ & 1 &  2 &  3 &  4 & 5 & $\vecu$  \\ \midrule
		\(p_{1}\) &		0.963888085 &		0.993552578 &		1.00351897 &		1.00056558 &		1.00116246 &		1 \\
		\(p_{2}\) &      1.24973 &              1.24584215 &        1.20049813 &        0.836871742 &      1.00020968 &        1 \\
		\(p_{3}\) &      2.49956544 &        2.49372368 &        2.47002669 &        2.47470852 &        1.999535 &            2 \\
		\(p_{4}\) &      1.25464633 &        1.31788044 &        1.14622359 &        1.00375959 &        0.999982026 &       1 \\
		\(p_{5}\) &      2.49514451 &        2.47823752 &        1.8914135 &          1.9937297 &          1.99998259 &        2 \\
		\(p_{6}\) &      0.970154676 &       0.999061501 &      0.999067816 &      1.00189432 &        1.00116117 &        1 \\
		\(p_{7}\) &      0.962706492 &       0.99653111 &        0.998111158 &      1.00094638 &        1.00215195 &        1 \\
		\(p_{8}\) &      1.24966644 &        0.880271698 &       0.999867849 &      0.999946039 &      0.999960889 &      1 \\
		\(p_{9}\) &      2.49946316 &        2.30207608 &        1.99927294 &        2.00013055 &        2.00010483 &        2 \\
		\(p_{10}\) &    0.990112197 &       0.999164353 &      0.999847143 &      0.999995161 &      0.999992747 &       1 \\
		\(p_{11}\) &    1.98133436 &        1.99986539 &        2.00026625 &        1.99985996 &        1.99988066 &        2 \\
		\(p_{12}\) &    0.964331475 &       0.997691152 &      0.998146016 &      1.00093643 &        1.00214616 &        1 \\
		\(p_{13}\) &    1.00174236 &        1.01281587 &        0.996999441 &       0.998250835 &      1.0001358 &          1 \\
		\(p_{14}\) &    1.24972392 &        0.87660879 &        1.00011661 &        1.00030994 &        0.999997189 &       1 \\
		\(p_{15}\) &    2.49955568 &        2.31341352 &        1.99944612 &        1.9989319 &          2.00004411 &        2 \\
		\(p_{16}\) &    1.00074035 &        1.00393178 &        0.99964541 &        0.99915064 &        1.00001059 &        1 \\
		\(p_{17}\) &    1.99600515 &        1.99083688 &        2.00061159 &        2.00146712 &        1.99998034 &        2 \\
		\(p_{18}\) &    1.00216469 &        1.00972037 &        0.997317303 &       0.999029072 &      1.00012768 &        1 \\
		\(p_{19}\) &    0.0919608293 &     0.100003407 &      0.100018715 &      0.0999866837 &     0.0999808376 &    0.1 \\
		\(p_{20}\) &    0.846854465 &       1.00004653 &        1.00005292 &        1.00012718 &        1.00005416 &        1 \\
		\(p_{21}\) &    0.0996315551 &     0.100000892 &      0.100015949 &      0.0999797001 &     0.0999778091 &    0.1 \\
		\(p_{22}\) &    0.0992356252 &     0.100017161 &      0.100015946 &      0.099996965 &       0.0999856293 &    0.1 \\
		\(p_{23}\) &    0.990213261 &       1.00001052 &        0.999978131 &      1.00000772 &        1.00005685 &        1 \\
		\(p_{24}\) &    0.0997384952 &     0.100016718 &      0.100017459 &      0.0999964905 &     0.0999821708 &    0.1 \\
		\(p_{25}\) &    0.0998074197 &     0.100021685 &      0.100026112 &      0.0999939244 &     0.100015632 &      0.1 \\
		\(p_{26}\) &    0.998740406 &       0.999963315 &      0.999962035 &      0.999931991 &      0.99991744 &        1 \\
		\(p_{27}\) &    0.09988352 &        0.10002433 &        0.100028713 &       0.0999987667 &    0.10002168 &        0.1 \\
		\(p_{28}\) &    1.02114856 &        1.00005354 &        0.999993562 &       0.999978013 &      0.999948959 &      1 \\
		\(p_{29}\) &    1.27248811 &        1.00181486 &        0.99953604 &        0.99994056 &        1.00039195 &        1 \\
		\(p_{30}\) &    1.26160143 &        1.00154535 &        0.998946553 &       1.00019337 &        1.001097 &           1 \\
		\(p_{31}\) &    1.00318475 &        1.00040209 &        0.999589148 &       1.00000564 &        1.00009912 &        1 \\
		\(p_{32}\) &    1.0213439 &          1.00144451 &        1.0003335 &          0.999927699 &      1.00017191 &        1 \\
		\(p_{33}\) &    1.02550256 &        0.99994568 &        1.00146268 &        1.00007653 &        1.00031567 &        1 \\
		\(p_{34}\) &    1.00842336 &        1.00032369 &        0.999865399 &       1.00002795 &        1.00022457 &        1 \\
		\(p_{35}\) &    1.02749726 &        1.00099571 &        1.00003443 &        0.999953653 &       1.00027132 &        1 \\
		\(p_{36}\) &    1.25837379 &        0.999210997 &       1.00039465 &        1.00009134 &        1.00017782 &        1 \\
		\bottomrule
	\end{tabular}
\end{adjustbox}
\caption{The evolution of the parameters vs. the number of data sets changing from 1 to 5, with 
\(t_0 = 0\), \(t_1 = 120\), \(\Delta t = 6.0\), 20 time measurements for each of the 8 components are given, \(\mathbf{u}_0 = 1.25\mathbf{u} \) and \(\alpha\) is determined using \eqref{eq:alphaformula}.}
\label{tab:tsmn_21_num_of_data_sets}
\end{table}

\begin{table}
\centering
\sisetup{
		round-mode = off,
		tight-spacing = true,
		table-number-alignment = center
}
\begin{tabularx}{\linewidth}{l@{\hskip .5in}S@{\hskip .75in}S}
	\toprule
		{Data sets} & {P} & {S} \\ \midrule
				1& 0.1&   	0.05\\
				2&0.1&   	0.13572\\
				3&0.1&   	0.3684\\
				4&0.1&   	1.0\\
				5&0.46416&   0.05\\
				6&0.46416&   0.13572\\
				7& 0.46416&   0.3684\\
				8&0.46416&   1.0\\
				9&2.1544&   0.05\\
				10&2.1544&   0.13572\\
				11&2.1544&   0.3684\\
				12& 2.1544&   1.0\\
				13& 10&   	0.05\\
				14& 10&   	0.13572\\
				15& 10&   	0.3684\\
				16& 10&   	1.0\\
		\bottomrule
	\end{tabularx}
\caption{The values of the control paramters \(P\) and \(S\) for estimations
that used 16 data sets.}
\label{tab:control_params_16}
\end{table}

\begin{table}
\centering
\sisetup{
		round-mode = off,
		tight-spacing = true,
		table-number-alignment = center
}
\begin{tabularx}{\linewidth}{l@{\hskip .5in}S@{\hskip .75in}S}
	\toprule
		{Data sets} & {P} & {S} \\ \midrule
				1				& 0.05	&	10\\
        2				& 0.3684 &  2.1544 \\
        3				& 1.0 		&0.1\\
        4				& 0.09286 &	2.1544\\
        5				& 0.13572 &	2.1544 \\
		\bottomrule
	\end{tabularx}
\caption{The values of the control paramters \(P\) and \(S\) for estimations
that used 5 data sets.}
\label{tab:control_params_05}
\end{table}

\end{document}